\documentstyle[10pt,epsf,epsfig,dp_delphititle]{dp_delphi}
\makeindex
\def\DpPaperGroup{PH-EP}
\def\DpPaperRef{2006-008}
\def\DpDate{9 March 2006}
\def\DpAuthors{DELPHI Collaboration}
\def\DpSubmit{(Accepted by Euro. Phys. J. C)}
\def\DpTitle{{Study of Triple-Gauge-Boson Couplings \boldmath $ZZZ$,
$ZZ\gamma$ and $Z\gamma\gamma$ at LEP}}
\def\DpComment{}
\def\DpEMail{}
\newcommand{\ra}{$\rightarrow$}
\newcommand{\mco}{\multicolumn} 
\newcommand {\ee} {\ifmmode \ e^+e^-  \else $e^+e^-$\fi }
\newcommand {\eeee} {\ifmmode \ e^+e^-e^+e^-  \else $e^+e^-e^+e^-$\fi }
\newcommand {\eemm} {\ifmmode \ e^+e^-\mu^+\mu^-  \else $e^+e^-\mu^+\mu^-$\fi}
\newcommand {\eett} {\ifmmode \ e^+e^-\tau^+\tau^-  \else $e^+e^-\tau^+\tau^-$\fi }
\newcommand {\mmmm} {\ifmmode \ \mu^+\mu^-\mu^+\mu^-  \else $\mu^+\mu^-\mu^+\mu^-$\fi}
\newcommand {\mmtt} {\ifmmode \ \mu^+\mu^-\tau^+\tau^-  \else $\mu^+\mu^-\tau^+\tau^-$\fi}
\newcommand {\tttt} {\ifmmode \ \tau^+\tau^-\tau^+tau^-  \else$ \tau^+\tau^-\tau^+\tau^-$\fi}
\newcommand {\eell} {\ifmmode \ e^+e^-l^+l^-  \else $e^+e^-l^+l^-$\fi}
\newcommand {\qqee} {\ifmmode \ q \bar q  e^+e^-  \else $q \bar q e^+e^-$\fi}
\newcommand {\qqmm} {\ifmmode \  q \bar q \mu^+\mu^-  \else $q \bar q \mu^+\mu^-$\fi}
\newcommand {\qqtt} {\ifmmode \  q \bar q \tau^+\tau^- \else $q \bar q \tau^+\tau^-$\fi}
\newcommand {\qqll} {\ifmmode \q \bar q  l^+l^-  \else $q \bar q l^+l^-$\fi}
\newcommand {\qqvv} {\ifmmode \ q \bar q  \nu \bar \nu \else $q \bar q \nu \bar  \nu$\fi}
\newcommand {\llvv} {\ifmmode \ l^+l^-  \nu \bar \nu \else $l^+ l^- \nu \bar \nu$\fi}
\newcommand {\eevv} {\ifmmode \ e^+e^-  \nu \bar \nu \else $e^+ e^- \nu \bar \nu$\fi}
\newcommand {\mmvv} {\ifmmode \  \mu^+\mu^- \nu \bar \nu \else 
$\mu^+\mu^- \nu \bar \nu$\fi}
\newcommand{\qqqq}{\ifmmode \ q {\overline q} q {\overline q}  \else $ q {\overline q} q {\overline q}$\fi}
\newcommand{\Wev}{\ifmmode \ W e \nu  \else $W e \nu$\fi}

\newcommand{\ZZ}{{\it ZZ}}
\newcommand{\eeZZ}{\mbox {\ee \ra\ \ZZ}}
\newcommand{\Zg}{{\it Z}$\gamma$}
\newcommand{\eeZg}{\mbox {\ee \ra\ \Zg}}
\newcommand{\Zgst}{{\it Z}$\gamma^*$}
\newcommand{\eeZgst}{\mbox {\ee \ra\ \Zgst}}
\newcommand{\ZZZ}{\ifmmode \ Z Z Z \else $ZZZ$\fi}
\newcommand{\VVV}{\ifmmode \ V_1^0 V_2^0 V_3^0 \else $V_1^0 V_2^0 V_3^0$\fi}
\newcommand{\ZZg}{\ifmmode \ Z Z \gamma \else $ZZ\gamma$\fi}
\newcommand{\Zgg}{\ifmmode \ Z \gamma \gamma \else $Z\gamma\gamma$\fi}
\newcommand{\WW}{\ifmmode \ W W  \else $WW$\fi}

\newcommand{\vvg}{\ifmmode \ \nu \overline{\nu} \gamma  \else \mbox {$\nu \overline{\nu} \gamma $}\fi}
\newcommand{\eevvg}{\mbox {$ \ee  \rightarrow \nu \overline{\nu} \gamma$}}
\newcommand{\qqg}{\ifmmode \ q \bar q  \gamma \else $q \bar q\gamma$\fi}
\newcommand{\qqbrg}{\ifmmode \ q \bar q  (\gamma) \else $q \bar q (\gamma)$\fi}
\newcommand{\eeqqg}{\mbox {$ \ee \rightarrow q \bar q \gamma$}}

\newcommand{\ffff} {\ifmmode \  f_1 \bar{f_2} f_3 \bar{f_4}  
                                                                 \else \mbox{$ f_1 \bar{f_2} f_3 \bar{f_4} $}\fi}

\newcommand{\ZVg}{ \ifmmode Z V \gamma \else \mbox{$ZV\gamma $}\fi}

\newcommand{\dgr}{\mbox{$^\circ$}}

\newcommand{\ba}{\begin{eqnarray}}
\newcommand{\ea}{\end{eqnarray}}

\begin{document}
%%%%%%%%%%%%%%%%%%%%%%%%%% They are a problem with Coll.Sty ?
\makeatletter
\makeatother
%%%%%%%%%%%%%%%%%%%%%%%%%% ??????????????????????????????????
%   Generate the title page
\begin{titlepage}
\pagenumbering{roman}

\CERNpreprint{\DpPaperGroup}{\DpPaperRef}   % Reference of the paper
\date{{\small\DpDate}}              % Date of the paper
\title{\DpTitle}                % Title of the paper
\address{\DpAuthors}                % General name of the author(s)

\begin{shortabs}                % Start the abstract
\noindent
%\input{abstract.tex}    % Paper abstract
%===================>  Abstract     =====> To be filled <=====%

Neutral triple-gauge-boson couplings \ZZZ, \ZZg\ and \Zgg\ have been studied with the DELPHI detector using data at energies between 183 and 208~GeV. Limits are derived on these couplings from an analysis of the reactions \eeZg, using data from the final states $\gamma f \bar f$, with $f$ = $q$ or $\nu$, from \eeZZ, using data from the four-fermion final states \qqqq, \qqmm, \qqee, \qqvv, \mmvv\ and \eevv, and from \eeZgst, in which the final state $\gamma$ is off mass-shell, using data from the four-fermion final states \qqee\ and \qqmm. No evidence for the presence of such couplings is observed, in agreement with the predictions of the Standard Model.

%=======================================================================%

\end{shortabs}

\vfill

\begin{center}
\DpSubmit \ \\      % Horrible hack to allow to have DpSubmit empty
\DpComment \ \\
\DpEMail \ \\
\end{center}

\vfill
\clearpage

\headsep 10.0pt

\addtolength{\textheight}{10mm}
\addtolength{\footskip}{-5mm}
\begingroup
%           Commands to process the author names
%
\newcommand{\DpName}[2]{\hbox{#1$^{\ref{#2}}$},\hfill}
\newcommand{\DpNameTwo}[3]{\hbox{#1$^{\ref{#2},\ref{#3}}$},\hfill}
\newcommand{\DpNameThree}[4]{\hbox{#1$^{\ref{#2},\ref{#3},\ref{#4}}$},\hfill}
\newskip\Bigfill \Bigfill = 0pt plus 1000fill
\newcommand{\DpNameLast}[2]{\hbox{#1$^{\ref{#2}}$}\hspace{\Bigfill}}

\small
%\footnotesize
\noindent
\DpName{J.Abdallah}{LPNHE}
\DpName{P.Abreu}{LIP}
\DpName{W.Adam}{VIENNA}
\DpName{P.Adzic}{DEMOKRITOS}
\DpName{T.Albrecht}{KARLSRUHE}
\DpName{R.Alemany-Fernandez}{CERN}
\DpName{T.Allmendinger}{KARLSRUHE}
\DpName{P.P.Allport}{LIVERPOOL}
\DpName{U.Amaldi}{MILANO2}
\DpName{N.Amapane}{TORINO}
% ...

\normalsize
\endgroup

\newpage
\titlefoot{Department of Physics and Astronomy, Iowa State
     University, Ames IA 50011-3160, USA
    \label{AMES}}
\titlefoot{IIHE, ULB-VUB,
     Pleinlaan 2, B-1050 Brussels, Belgium
    \label{BRUSSELS}}
\titlefoot{Physics Laboratory, University of Athens, Solonos Str.
     104, GR-10680 Athens, Greece
    \label{ATHENS}}
\titlefoot{Department of Physics, University of Bergen,
     All\'egaten 55, NO-5007 Bergen, Norway
    \label{BERGEN}}
\titlefoot{Dipartimento di Fisica, Universit\`a di Bologna and INFN,
     Via Irnerio 46, IT-40126 Bologna, Italy
    \label{BOLOGNA}}
\titlefoot{Centro Brasileiro de Pesquisas F\'{\i}sicas, rua Xavier Sigaud 150,
     BR-22290 Rio de Janeiro, Brazil
    \label{BRASIL-CBPF}}
\titlefoot{Inst. de F\'{\i}sica, Univ. Estadual do Rio de Janeiro,
     rua S\~{a}o Francisco Xavier 524, Rio de Janeiro, Brazil
    \label{BRASIL-IFUERJ}}
\titlefoot{Coll\`ege de France, Lab. de Physique Corpusculaire, IN2P3-CNRS,
     FR-75231 Paris Cedex 05, France
    \label{CDF}}
\titlefoot{CERN, CH-1211 Geneva 23, Switzerland
    \label{CERN}}
\titlefoot{Institut de Recherches Subatomiques, IN2P3 - CNRS/ULP - BP20,
     FR-67037 Strasbourg Cedex, France
    \label{CRN}}
\titlefoot{Now at DESY-Zeuthen, Platanenallee 6, D-15735 Zeuthen, Germany
    \label{DESY}}
\titlefoot{Institute of Nuclear Physics, N.C.S.R. Demokritos,
     P.O. Box 60228, GR-15310 Athens, Greece
    \label{DEMOKRITOS}}
\titlefoot{FZU, Inst. of Phys. of the C.A.S. High Energy Physics Division,
     Na Slovance 2, CZ-182 21, Praha 8, Czech Republic
    \label{FZU}}
\titlefoot{Dipartimento di Fisica, Universit\`a di Genova and INFN,
     Via Dodecaneso 33, IT-16146 Genova, Italy
    \label{GENOVA}}
\titlefoot{Institut des Sciences Nucl\'eaires, IN2P3-CNRS, Universit\'e
     de Grenoble 1, FR-38026 Grenoble Cedex, France
    \label{GRENOBLE}}
\titlefoot{Helsinki Institute of Physics and Department of Physical Sciences,
     P.O. Box 64, FIN-00014 University of Helsinki, 
     \indent~~Finland
    \label{HELSINKI}}
\titlefoot{Joint Institute for Nuclear Research, Dubna, Head Post
     Office, P.O. Box 79, RU-101 000 Moscow, Russian Federation
    \label{JINR}}
\titlefoot{Institut f\"ur Experimentelle Kernphysik,
     Universit\"at Karlsruhe, Postfach 6980, DE-76128 Karlsruhe,
     Germany
    \label{KARLSRUHE}}
\titlefoot{Institute of Nuclear Physics PAN,Ul. Radzikowskiego 152,
     PL-31142 Krakow, Poland
    \label{KRAKOW1}}
\titlefoot{Faculty of Physics and Nuclear Techniques, University of Mining
     and Metallurgy, PL-30055 Krakow, Poland
    \label{KRAKOW2}}
\titlefoot{Universit\'e de Paris-Sud, Lab. de l'Acc\'el\'erateur
     Lin\'eaire, IN2P3-CNRS, B\^{a}t. 200, FR-91405 Orsay Cedex, France
    \label{LAL}}
\titlefoot{School of Physics and Chemistry, University of Lancaster,
     Lancaster LA1 4YB, UK
    \label{LANCASTER}}
\titlefoot{LIP, IST, FCUL - Av. Elias Garcia, 14-$1^{o}$,
     PT-1000 Lisboa Codex, Portugal
    \label{LIP}}
\titlefoot{Department of Physics, University of Liverpool, P.O.
     Box 147, Liverpool L69 3BX, UK
    \label{LIVERPOOL}}
\titlefoot{Dept. of Physics and Astronomy, Kelvin Building,
     University of Glasgow, Glasgow G12 8QQ
    \label{GLASGOW}}
\titlefoot{LPNHE, IN2P3-CNRS, Univ.~Paris VI et VII, Tour 33 (RdC),
     4 place Jussieu, FR-75252 Paris Cedex 05, France
    \label{LPNHE}}
\titlefoot{Department of Physics, University of Lund,
     S\"olvegatan 14, SE-223 63 Lund, Sweden
    \label{LUND}}
\titlefoot{Universit\'e Claude Bernard de Lyon, IPNL, IN2P3-CNRS,
     FR-69622 Villeurbanne Cedex, France
    \label{LYON}}
\titlefoot{Dipartimento di Fisica, Universit\`a di Milano and INFN-MILANO,
     Via Celoria 16, IT-20133 Milan, Italy
    \label{MILANO}}
\titlefoot{Dipartimento di Fisica, Univ. di Milano-Bicocca and
     INFN-MILANO, Piazza della Scienza 3, IT-20126 Milan, Italy
    \label{MILANO2}}
\titlefoot{IPNP of MFF, Charles Univ., Areal MFF,
     V Holesovickach 2, CZ-180 00, Praha 8, Czech Republic
    \label{NC}}
\titlefoot{NIKHEF, Postbus 41882, NL-1009 DB
     Amsterdam, The Netherlands
    \label{NIKHEF}}
\titlefoot{National Technical University, Physics Department,
     Zografou Campus, GR-15773 Athens, Greece
    \label{NTU-ATHENS}}
\titlefoot{Physics Department, University of Oslo, Blindern,
     NO-0316 Oslo, Norway
    \label{OSLO}}
\titlefoot{Dpto. Fisica, Univ. Oviedo, Avda. Calvo Sotelo
     s/n, ES-33007 Oviedo, Spain
    \label{OVIEDO}}
\titlefoot{Department of Physics, University of Oxford,
     Keble Road, Oxford OX1 3RH, UK
    \label{OXFORD}}
\titlefoot{Dipartimento di Fisica, Universit\`a di Padova and
     INFN, Via Marzolo 8, IT-35131 Padua, Italy
    \label{PADOVA}}
\titlefoot{Rutherford Appleton Laboratory, Chilton, Didcot
     OX11 OQX, UK
    \label{RAL}}
\titlefoot{Dipartimento di Fisica, Universit\`a di Roma II and
     INFN, Tor Vergata, IT-00173 Rome, Italy
    \label{ROMA2}}
\titlefoot{Dipartimento di Fisica, Universit\`a di Roma III and
     INFN, Via della Vasca Navale 84, IT-00146 Rome, Italy
    \label{ROMA3}}
\titlefoot{DAPNIA/Service de Physique des Particules,
     CEA-Saclay, FR-91191 Gif-sur-Yvette Cedex, France
    \label{SACLAY}}
\titlefoot{Instituto de Fisica de Cantabria (CSIC-UC), Avda.
     los Castros s/n, ES-39006 Santander, Spain
    \label{SANTANDER}}
\titlefoot{Inst. for High Energy Physics, Serpukov
     P.O. Box 35, Protvino, (Moscow Region), Russian Federation
    \label{SERPUKHOV}}
\titlefoot{J. Stefan Institute, Jamova 39, SI-1000 Ljubljana, Slovenia
    \label{SLOVENIJA1}}
\titlefoot{Laboratory for Astroparticle Physics,
     University of Nova Gorica, Kostanjeviska 16a, SI-5000 Nova Gorica, Slovenia
    \label{SLOVENIJA2}}
\titlefoot{Department of Physics, University of Ljubljana,
     SI-1000 Ljubljana, Slovenia
    \label{SLOVENIJA3}}
\titlefoot{Fysikum, Stockholm University,
     Box 6730, SE-113 85 Stockholm, Sweden
    \label{STOCKHOLM}}
\titlefoot{Dipartimento di Fisica Sperimentale, Universit\`a di
     Torino and INFN, Via P. Giuria 1, IT-10125 Turin, Italy
    \label{TORINO}}
\titlefoot{INFN,Sezione di Torino and Dipartimento di Fisica Teorica,
     Universit\`a di Torino, Via Giuria 1,
     IT-10125 Turin, Italy
    \label{TORINOTH}}
\titlefoot{Dipartimento di Fisica, Universit\`a di Trieste and
     INFN, Via A. Valerio 2, IT-34127 Trieste, Italy
    \label{TRIESTE}}
\titlefoot{Istituto di Fisica, Universit\`a di Udine and INFN,
     IT-33100 Udine, Italy
    \label{UDINE}}
\titlefoot{Univ. Federal do Rio de Janeiro, C.P. 68528
     Cidade Univ., Ilha do Fund\~ao
     BR-21945-970 Rio de Janeiro, Brazil
    \label{UFRJ}}
\titlefoot{Department of Radiation Sciences, University of
     Uppsala, P.O. Box 535, SE-751 21 Uppsala, Sweden
    \label{UPPSALA}}
\titlefoot{IFIC, Valencia-CSIC, and D.F.A.M.N., U. de Valencia,
     Avda. Dr. Moliner 50, ES-46100 Burjassot (Valencia), Spain
    \label{VALENCIA}}
\titlefoot{Institut f\"ur Hochenergiephysik, \"Osterr. Akad.
     d. Wissensch., Nikolsdorfergasse 18, AT-1050 Vienna, Austria
    \label{VIENNA}}
\titlefoot{Inst. Nuclear Studies and University of Warsaw, Ul.
     Hoza 69, PL-00681 Warsaw, Poland
    \label{WARSZAWA}}
\titlefoot{Now at University of Warwick, Coventry CV4 7AL, UK
    \label{WARWICK}}
\titlefoot{Fachbereich Physik, University of Wuppertal, Postfach
     100 127, DE-42097 Wuppertal, Germany \\
\noindent
{$^\dagger$~deceased}
    \label{WUPPERTAL}}

\addtolength{\textheight}{-10mm}
\addtolength{\footskip}{5mm}
\clearpage

\headsep 30.0pt
\end{titlepage}

%%%%%%%%%%%%%%%%%%%%%%%%%
%   Change for the document body
%%\pagestyle{heading}                   % for page numbering
\pagenumbering{arabic}                  % page numbering in number
\setcounter{footnote}{0}                %
\large

%%%%%%%%%%%%%%%%%%%%%%%%%%%%%
%% Switch on/off linenumbers
%\linenumbers 
%%%%%%%%%%%%%%%%%%%%%%%%%%%%%

%\input{document}      
%#######################################################################%

\section{Introduction}
\label{sec:intro}
One of the important properties of the Standard Model which can be tested at LEP2 is its non-Abelian character, leading to the prediction of triple-gauge-boson couplings. However, while non-zero values of these couplings are predicted for the charged ($WW\gamma$, $WWZ$) sector, the $SU(2) \times U(1)$ symmetry of the Standard Model predicts the absence of such couplings in the neutral sector, namely at the \ZZZ, \ZZg\ and \Zgg\ vertices. This paper describes an investigation of this prediction by DELPHI using LEP2 data taken between 1997 and 2000 at energies between 183 and 208~GeV. 

\subsection{Phenomenology of the neutral triple-gauge-boson vertex}
\label{sec:intro_phen}

\begin{figure}[h]
   \centerline{
   \epsfig{file=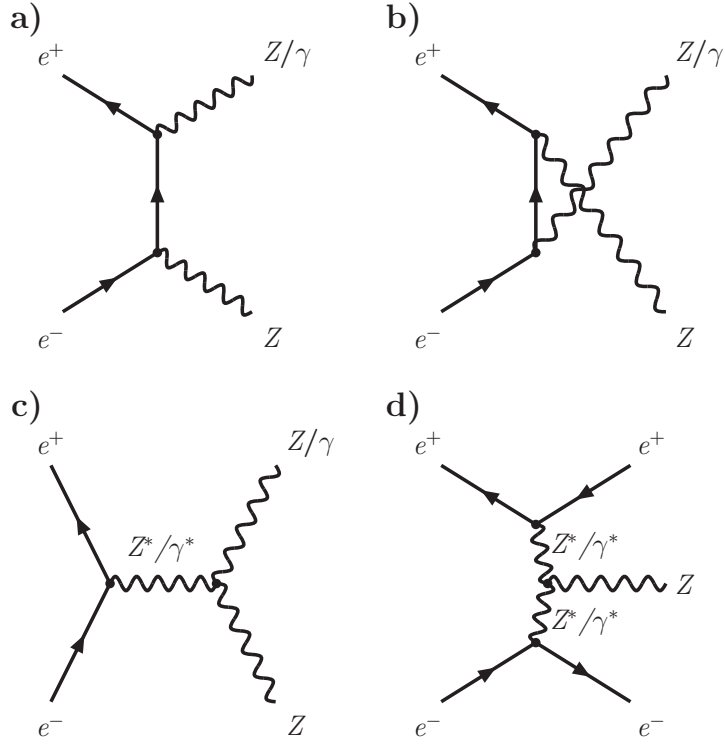,width=9.5cm}}
\caption{a), b) Lowest order Feynman diagrams for the production of two gauge bosons \ZZ\ and \Zg\ by Standard Model processes, where both on- and off-shell $\gamma$ production is implied. 
c) Production of the same final states via an anomalous interaction among three neutral gauge bosons.
d) Production of the $Z$\ee\ final state via an anomalous neutral triple-gauge-boson coupling. }
\label{fig:diagrams} 
\end{figure}

Within the Standard Model, production of two neutral gauge bosons in \ee\ collisions proceeds at lowest order via the $t$- or $u$-channel exchange of an electron. These processes are shown in figures~\ref{fig:diagrams}a) and~b), where both on- and off-shell $\gamma$ production is implied, as is the subsequent decay of the final state $Z$ or off-shell $\gamma$ into a fermion-antifermion pair. Figure~\ref{fig:diagrams}c) shows a contribution to production of the same final states which could come from physics beyond the Standard Model by the $s$-channel exchange of a virtual $\gamma$ or $Z$ via a neutral triple-gauge-boson coupling. In the reactions \eeZg\ and \eeZZ\ the final state can, to a good approximation, be considered to be of two on-shell bosons, so that only the exchanged boson at the triple-gauge-boson vertex need necessarily be considered as off-shell, while in the reaction \eeZgst\ both the exchanged boson and the outgoing $\gamma^*$ are off-shell.\footnote{Throughout this paper, we write ``$V^*$" when we wish to be explicit that a vector
boson $V$ is off mass-shell. When it is clear that it is on mass-shell, or when it can be either on or off mass-shell, the star (``$^*$") is omitted.} A further process containing a neutral triple-gauge-boson coupling with two of the bosons off-shell is shown in figure~\ref{fig:diagrams}d); here a single $Z$ is produced in the final state $Z$\ee\ via fusion of two exchanged vector bosons.

The phenomenology of the case where two of the three neutral gauge bosons interacting at the \VVV\ vertex are on mass-shell has been described in~\cite{hagiwara}. In this case, there are twelve independent anomalous couplings satisfying Lorentz invariance and Bose symmetry. Calling $V$ the exchanged boson ($V = Z,\gamma$), the couplings $f_i^V$ ($i=$4,5) produce a \ZZ\ final state and $h_i^V$ ($i=1\cdots 4$) the {\it Z}$\gamma$ final state. The couplings $f^{V}_{5}$, $h^{V}_{3}$ and $h^{V}_{4}$ are CP-conserving and $f^{V}_{4}$, $h^{V}_{1}$ and $h^{V}_{2}$ are CP-violating. There are no couplings common to production of both the \ZZ\ and \Zg\ final states. 

A complete phenomenological description of the anomalous neutral gauge couplings in the case where one, two or three of the gauge bosons interacting at the \VVV\ vertex may be off mass-shell has been developed in~\cite{renard}. Following the treatment of the charged 
triple-gauge-boson vertex developed, for instance, in~\cite{bilenky,yb}, all the Lorentz-invariant forms which can contribute to the \ZZZ, \ZZg\ and \Zgg\  vertices are listed, imposing Bose symmetry as appropriate. An effective Lagrangian model is then developed in terms of the operators of lowest dimension which are required to reconstruct fully all the vertex forms, and which affect only the neutral triple-gauge-boson vertex.\footnote{The \VVV\ vertex functions receive contributions from both transverse and scalar terms, the latter contributing in the case where one off-shell $Z$ decays to a heavy fermion pair through its axial coupling. In the analysis of LEP data only transverse terms need be considered, due to the negligible contribution of $Z$ \ra $t \bar t$ decays. The contribution of scalar terms is therefore ignored in the following.} This leads to a Lagrangian with operators of dimension, $d$, ranging from $d=6$ to $d=12$. 
Such an expansion is valid in the case where the new physics energy scale, $\Lambda$, represented by the operators is very high, at least satisfying the condition $\Lambda \gg (m_Z, \sqrt{s})$, and the relative contribution from an operator of dimension $d$ may be expected to be suppressed by a factor $1/\Lambda^{(d-4)}$. In the analysis we report here, we have considered only the lowest dimension operators contributing to the parameters we have determined. In addition to satisfying Lorentz and Bose symmetry, the operators are required to be $U(1)_{em}$-invariant, and both $CP$-conserving operators, ${\cal O}$, and $CP$-violating operators, $\tilde{{\cal O}}$, with coefficients $\ell$ and $\tilde{\ell}$, respectively, are considered: 

\begin{equation}
{\cal L} = e (\sum_{i,C\!P+} \ell_i^{V_1^0 V_2^0 V_3^0} {\cal O}_i^{V_1^0 V_2^0 V_3^0}
    + \sum_{i,C\!P-} \tilde{\ell}_i^{V_1^0 V_2^0 V_3^0} \tilde{{\cal O}}_i^{V_1^0 V_2^0 V_3^0})  \, .
\label{eq:lagrangian}
\end{equation}
 
\noindent Of the operators included in the sum defined above, some affect the $V^0 ZZ$ and $V^0 Z \gamma^*$ vertices ($V^0 \equiv Z,\gamma$), some the $V^0 Z \gamma^*$ vertex only, and some the $V^0 Z \gamma^*$ and $V^0 Z \gamma$ vertices; none contribute to all three vertices. In~\cite{renard} a connection is made between the coefficients $\ell_i$ and $\tilde{\ell}_i$ of the operators in the effective Lagrangian describing the general \VVV\ vertex and the dimensionless coefficients $h$ and $f$ describing on-shell \Zg\ and \ZZ\ production, respectively: retaining only the terms corresponding to contributions from operators of lowest dimension, each of the coefficients  $h_1^V$, $h_3^V$, $f_4^V$ and $f_5^V$ (which are dimensionless) is related to one operator of dimension $d= 6$ by $f, h = \ell^{V_1^0 V_2^0 V_3^0} m_Z^2$. The lowest dimension operators contributing to $h_2^V$ and $h_4^V$ have $d=8$. 

As in the case of the charged triple-gauge-boson couplings, a further simplification in the possible structure of the effective Lagrangian is obtained by imposition of $SU(2) \times U(1)$ invariance on its form. Such a form is presented in an addendum~\cite{renard-addendum} to~\cite{renard}, and the effective Lagrangian reduces to a sum of two terms, both with dimension $d=8$, one (${\cal O}_{SU(2) \times U(1)}$) $CP$-conserving and one ($\tilde{{\cal O}}_{SU(2) \times U(1)}$) $CP$-violating. This simplification leads to constraints between some of the $\ell_i^{\VVV}$ or $\tilde{\ell}_i^{\VVV}$ defined in equation~(\ref{eq:lagrangian}):

\begin{equation}
 \ell_1^{\ZZZ} \cot \theta_W = \ell_1^{\ZZg} = - \ell_2^{\ZZg} =- \ell_1^{\Zgg} \tan \theta_W
 = \frac{v^2}{4} \ell_{SU(2) \times U(1)}   \ , 
\label{eq:G1}
\end{equation}
\begin{equation}
  \tilde{\ell}_1^{\ZZZ} \cot \theta_W = \tilde{\ell}_1^{\ZZg} = -\tilde{\ell}_3^{\ZZg} = 
            - \tilde{\ell}_1^{\Zgg} \tan \theta_W  =  \frac{v^2}{4} \tilde{\ell}_{SU(2) \times U(1)}  \  ,
\label{eq:G2}
\end{equation}

\noindent where $\theta_W$ is the weak mixing angle, $v$ is the vacuum expectation value of the Higgs field and $\ell_{SU(2) \times U(1)}$, $ \tilde{\ell}_{SU(2) \times U(1)}$ are the coefficients of the operators ${\cal O}_{SU(2) \times U(1)}$ and $\tilde{{\cal O}}_{SU(2) \times U(1)}$, respectively. If applied solely to the on-shell channels \Zg\ and \ZZ, these conditions become, respectively:

\begin{equation}
 f_5^Z \cot \theta_W = h_3^Z = -f_5^\gamma  = -h_3^\gamma \tan \theta_W 
  = m_Z^2   \frac{v^2}{4} \ell_{SU(2) \times U(1)}   \ ,
 \label{eq:A1} 
\end{equation} 
\begin{equation}
 f_4^Z \cot \theta_W = h_1^Z = -f_4^\gamma   = -h_1^\gamma \tan \theta_W 
   = m_Z^2   \frac{v^2}{4} \tilde{\ell}_{SU(2) \times U(1)}  \ .
\label{eq:A2}
\end{equation} 

\noindent The $SU(2) \times U(1)$-conserving Lagrangian considered in~\cite{renard} is constructed so as to affect only the neutral gauge boson and Higgs sectors, and an alternative form, which would additionally affect off-mass-shell charged gauge boson production, has been proposed in~\cite{alcaraz}. This leads to a Lagrangian with four possible terms, two $CP$-conserving (${\cal O}_8^A$, ${\cal O}_8^B$) and two $CP$-violating ($\tilde{{\cal O}}_8^A$, $\tilde{{\cal O}}_8^B$) and hence to looser constraints between the possible contributing operators: in each of the sets of conditions~(\ref{eq:G1}) -~(\ref{eq:A2}) listed above, the relations corresponding in the diagrams of figure~\ref{fig:diagrams}  to $\gamma$ and $Z$ exchange decouple, giving, for instance in the case of~(\ref{eq:G1}), the separate conditions

\begin{equation}
 \ell_1^{\ZZZ} \cot \theta_W = \ell_1^{\ZZg}  \ \mathrm{and} \  \ell_2^{\ZZg} 
        = \ell_1^{\Zgg} \tan \theta_W \ , 
\label{eq:Q1}
\end{equation}

\noindent with an analogous separation in conditions~(\ref{eq:G2}) -~(\ref{eq:A2}). This leads to four coefficients, $\ell_8^{A,B}$, $\tilde{\ell}_8^{A,B}$, related to the respective Lagrangian operators by appropriate factors of $m_Z$ and $v$. In both the stronger and weaker of these sets of constraints (which we refer to as the Gounaris-Layssac-Renard and Alcaraz constraints, with respect to the authorship of references~\cite{renard} and~\cite{alcaraz}), the gauge-invariant operators all now contribute to all three neutral triple-gauge-boson vertices, $V^0 ZZ$, $V^0 Z \gamma^*$ and $V^0 Z \gamma$. 

In order to study the \VVV\ vertex, three physical final states have been defined from the data: \Zg, \Zgst\ and \ZZ. The first of these is a three-body final state comprising the $Z$ decay products and a detected photon, while the other two are four-fermion final states with, respectively, one or two fermion-antifermion pairs having mass in the $Z$ region. Given the phenomenology summarized above, we have then chosen to determine the following parameters in our study:
\begin{itemize}
\item{Using data from the final states \ZZ\ and \Zgst, values are determined for the coefficients of each of the four $d=6$ operators which are related in the on-shell limit to one of the $f$ coefficients defined in the on-shell formalism of reference~\cite{hagiwara}. Similarly, using data from the final states \Zg, \Zgst\ and \ZZ, values of the coefficients of the four $d=6$ operators related to the on-shell $h$ coefficients are determined. In the latter case, the \ZZ\ data are used as well as the \Zgst, as the off-shell $\gamma$ couples to the $f \bar f$ system over the whole of the four-fermion phase space. However, in these studies the statistical contribution of the off-shell final states compared to that of on-shell \Zg\ or \ZZ\ production is very small, so that the values of the parameters determined, quoted in dimensionless form, $\ell^{V_1^0 V_2^0 V_3^0} m_Z^2$, are directly comparable with published results using data from on-shell channels, and the relevant respective likelihood distributions may be combined.}

\item{The $V^0 Z \gamma^*$ vertex is studied on its own by determining the coefficients of the lowest dimension operators which affect solely these vertices. There are two such operators, both of dimension $d=8$, one $CP$-conserving and involving $s$-channel $\gamma$ exchange in the production process illustrated in figure~\ref{fig:diagrams}c), and the other $CP$-violating and involving $s$-channel $Z$ exchange in the same diagram. Again, data from both the \Zgst\ and \ZZ\ final states were used in the determination of the coefficients of these operators, and the coefficients are quoted in dimensionless form: $\ell^{V_1^0 V_2^0 V_3^0} m_Z^4$.}
\item{The coefficients of the $SU(2) \times U(1)$-conserving operators are determined, using both the Gounaris-Layssac-Renard and the Alcaraz constraints. They are quoted in a dimensionless form, such that in the on-shell limit they become equal to one of the $h_i^V$ occurring in the constraint equations~(\ref{eq:A1}) and~(\ref{eq:A2}) above. }
\end{itemize}
\noindent A list of the parameters we have determined, the definitions of the operators to which they refer and (where relevant) the on-shell coefficients to which they are related is given in table~\ref{table:parameters}.

\begin{table}
\begin{center}
\begin{tabular}{|c|c|c|c|}
\hline 
Vertices & Parameter             & Lagrangian  &  Related on-shell    \\ 
 affected&                       &   Operator  &  coefficient         \\ \hline
\mco{4}{|l|}{a)}                                           \\ \hline \hline
\ZZZ         &$\tilde{\ell}_1^{\ZZZ} m_Z^2$ 
                                 &   $-Z_\sigma (\partial^\sigma Z_\nu ) (\partial_\mu Z^{\mu \nu})$                    & $f_4^Z$               \\
             &$\ell_1^{\ZZZ} m_Z^2$   
                                 & $\tilde{Z}_{\mu \nu} ( \partial_\sigma Z^{\sigma \mu} ) Z^\nu$
                                              &  $f_5^Z$            \\ \hline
\ZZg         &$\tilde{\ell}_3^{\ZZg} m_Z^2$  
                                & $-(\partial_\mu F^{\mu \beta}) Z_\alpha (\partial^\alpha Z_\beta )  $                 & $f_4^\gamma$     \\
             &$\ell_2^{\ZZg} m_Z^2 $             
                                & $\tilde{Z}^{\mu \nu} Z_\nu (\partial^\sigma F_{\sigma \mu} )  $         & $f_5^\gamma$     \\
             &$\tilde{\ell}_1^{\ZZg} m_Z^2$  
                               & $-F^{\mu \beta} Z_\beta (\partial^\sigma Z_{\sigma \mu} ) $                           & $h_1^Z$           \\
             &$\ell_1^{\ZZg} m_Z^2 $             
                               & $-\tilde{F}_{\mu \nu} Z^\nu (\partial_\sigma Z^{\sigma \mu} )  $         & $h_3^Z$            \\ \hline
\Zgg        &$\tilde{\ell}_1^{\Zgg} m_Z^2$   
                               & $-(\partial^\sigma F_{\sigma \mu}) Z_\beta F^{\mu \beta}$                       & $h_1^\gamma$ \\  
            &$\ell_1^{\Zgg } m_Z^2$             
                               & $-\tilde{F}_{\rho \alpha} (\partial_\sigma F^{\sigma \rho}) Z^\alpha$  & $h_3^\gamma$ \\ \hline \hline 
\mco{4}{|l|}{b)}                                               \\ \hline
\ZZg        &$\tilde{\ell}_4^{\ZZg}  m_Z^4$   
                                & $\partial^\mu F_{\mu\nu} (\Box \partial^\nu Z_\alpha) Z^\alpha$             &        --       \\ \hline
\Zgg        &$\ell_2^{\Zgg} m_Z^4$               
                                & $\Box \tilde{F}^{\mu\nu} (\partial^\sigma  F_{\sigma\mu}) Z_\nu$      &        -- \\ \hline \hline
\mco{4}{|l|}{c)\ i)}                                     \\ \hline
\ZZZ\ \ZZg\ \Zgg
            &$- \cot \theta_W m_Z^2 \frac{v^2}{4} \tilde{\ell}_{SU(2) \times U(1)}$                    &  $ i B_{\mu\nu} (\partial_\sigma B^{\sigma\mu}) (\Phi^\dagger D^\nu \Phi)$    &  $h_1^\gamma$     \\
            &$- \cot \theta_W m_Z^2 \frac{v^2}{4} \ell_{SU(2) \times U(1)}$                           & $ i \tilde{B}_{\mu\nu} (\partial_\sigma B^{\sigma\mu}) (\Phi^\dagger D^\nu \Phi)  $   
                                             &  $h_3^\gamma$ \\   \hline
\mco{4}{|l|}{\ \ \ ii)}                                            \\ \hline
\ZZg        &$- \cot \theta_W m_Z^2 \frac{v^2}{4} \tilde{\ell}_8^A$  
                                &  $ i B_{\mu\nu} (\partial_\sigma B^{\sigma\mu}) (\Phi^\dagger D^\nu \Phi) $   &  $h_1^\gamma$      \\
            &$- \cot \theta_W m_Z^2 \frac{v^2}{4} \ell_8^A$  
                                & $ i \tilde{B}_{\mu\nu} (\partial_\sigma B^{\sigma\mu}) (\Phi^\dagger D^\nu \Phi) $   &  $h_3^\gamma$  \\ \hline
\ZZZ\ \Zgg  &$- \cot \theta_W m_Z^2 \frac{v^2}{4} \tilde{\ell}_8^B$  
                                &  $ i B_{\mu\nu} (\partial_\sigma W_I^{\sigma\mu}) (\Phi^\dagger \tau_I D^\nu \Phi) $ 
                                              &  $h_1^Z$  \\
            &$- \cot \theta_W m_Z^2 \frac{v^2}{4} \ell_8^B$ 
                                &  $ i \tilde{B}_{\mu\nu} (\partial_\sigma W_I^{\sigma\mu}) (\Phi^\dagger \tau_I D^\nu \Phi) $ 
                                              & $h_3^Z$ \\  \hline  
\end{tabular}
\caption{
Parameters determined in this study, corresponding Lagrangian operators in the models of references~\cite{renard} and~\cite{alcaraz}, and (where appropriate) related on-shell parameters: a)~Coefficients of lowest dimension operators contributing to \ZZ\ and \Zgst\ production or to \Zgst\ and \Zg\ production; b)~Coefficients of lowest dimension operators affecting only the $V^0 Z \gamma^*$ vertices; c)~Coefficients of $SU(2) \times U(1)$-conserving operators according to i) the Gounaris-Layssac-Renard constraints and ii) the Alcaraz constraints. The constraints are given in the text. The vertices $V_1^0 V_2^0 V_3^0$ affected by these operators (without distinguishing the $V_i^0$ as on- or off-mass-shell) are indicated in column 1. The fields $Z_\mu$, $F_\mu$, $B_\mu$ and $W_\mu$ represent the $Z$, photon, $U(1)_Y$ and $SU(2)_L$ fields, respectively; $\tilde{Z}_{\mu\nu}$, $\tilde{F}_{\mu\nu}$ and $\tilde{B}_{\mu\nu}$ are the contractions of the respective field tensors with the four-dimensional antisymmetric tensor; $\Phi$ is the Higgs field and $v$ its vacuum expectation value,  $D$ represents the covariant derivative of $SU(2) \times U(1)$, and $\tau_I$ are the Pauli matrices.
}
\label{table:parameters}
\end{center}
\end{table}

\subsection{Experimental considerations}
\label{sec:intro_exp}

Of the three final state channels, \Zg, \ZZ\ and \Zgst, defined in the previous section, the most precise limits on anomalous couplings are derived from the first, when the final state photon is on-shell. In this channel, the kinematic region with high photon energy and large photon polar angle is most sensitive to the anomalous couplings, and in this region the anomalous interactions give rise to a change in the total rate and to an enhancement of the production of longitudinally polarized $Z$ bosons. Our analysis covers two reactions to which the diagrams describing \Zg\ production provide the dominant  contribution: \eevvg, in which the observed number of events is compared with the number predicted from the total cross-section for this process, and \eeqqg, with the $q \bar{q}$ system coming predominantly from $Z$ decay, in which the distribution of the decay angle of the $Z$ in its rest frame with respect to the direction of the $Z$ in the overall centre of mass is studied. The present analysis uses data from LEP2 at energies ranging from 189 to 208~GeV. Previous DELPHI results on this channel can be found in~\cite{zg}; they used data with energies up to $\sqrt{s}$ = 172~GeV, and the limits were obtained using an analysis based only on the value of the observed total cross-section. 

The total \ZZ\ cross-section is also sensitive to the anomalous couplings, and the sensitivity increases strongly with $\sqrt{s}$. Large interference between Standard Model and anomalous amplitudes arises for $CP$-conserving couplings (especially for $f_5^Z$) when considering the differential cross-section $d\sigma/d |\cos\theta_Z |$, where $\theta_Z$ is the $Z$ production angle with respect to the beam axis. The analysis reported here is based on a study of this differential distribution in the LEP2 data in the energy range 183 to 208~GeV. DELPHI has previously reported a study of the  \ZZ\ production cross-section in all visible \ffff\ final states in these data~\cite{zz}. The same sets of identified events have been used in the present analysis, with the exception of the \qqtt, $\tau^+ \tau^- \nu \bar \nu$ and $l^+l^-l^+l^-$ channels, which are not used. 

In a separate publication~\cite{zgst}, DELPHI has studied the \Zgst\ final state in the same LEP2 data as used for the channels described above, reporting on a comparison of the cross-section for \Zgst\ production in various channels with Standard Model predictions. We use the samples identified in~\cite{zgst} in the \qqee\ and \qqmm\ final states in the present analysis, which thus represents an interpretation of these data for the first time in terms of possible anomalous gauge couplings. The data were examined as a function of the bidimensional ($M_{l^+ l^-}$,  $M_{q \bar q}$) mass distribution, requiring one of them to be in the region of the $Z$ mass, and they were also divided into two regions of the $l^+ l^-$ polar angle with respect to the beam direction (equivalent to the variable $\theta_Z$ used in the analysis of \ZZ\ events). 

Limits on anomalous neutral gauge couplings in the \Zg\ and \ZZ\ final states have been reported by other LEP experiments; recent published results may be found in the papers listed in~\cite{opal1,lep_ntgc}. 

\section{Experimental details and analysis}
\label{sec:exp}

Events were recorded in the DELPHI detector. Detailed descriptions of the DELPHI components can be found in~\cite{delphi_det} and the description of its performance and of the luminosity monitor can be found in~\cite{delphi_perf}. The trigger system is described in~\cite{trigger}. For LEP2 operations, the vertex detector was upgraded~\cite{delphi_det2}, and a set of scintillation counters was added to veto photons in the blind regions of the electromagnetic calorimetry at polar angles around $\theta = 40^\circ$. The performance of the detector was simulated using the program DELSIM~\cite{delphi_perf}, which was interfaced to the programs used in the generation of Monte Carlo events and to the programs used to simulate the hadronization of quarks from $Z$ and $\gamma^\ast$ decay or from background processes. During the year 2000, one sector (1/12) of the Time Projection Chamber (TPC), DELPHI's main tracking device, was inactive for about a quarter of the data-taking period. The effect of this was taken into account in the detector simulation and in the determination of cross-sections from the data.

The selection of events in the three physical final states, \Zg, \ZZ\ and \Zgst, considered in this paper, and the simulation of the processes contributing to signals and backgrounds are described in the following subsections. In the case of the \ZZ\ and \Zgst\ samples, the reader is referred to recent DELPHI publications on the production of these final states (references~\cite{zz,zgst}, respectively) for a full description of the event selection procedures. The event samples used in the present analysis of these two final states have been selected using essentially identical procedures to those described in~\cite{zz,zgst}, and cover the same energy range (183 - 208~GeV). These procedures are summarized, respectively, in sections~\ref{sec:exp_zz} and~\ref{sec:exp_zgst} below, and any changes from the methods described in~\cite{zz,zgst} are mentioned. DELPHI has also reported a study of events observed at LEP2 in which only photons and missing energy were detected~\cite{gx}. The present analysis uses data in the part of the kinematic region covered in~\cite{gx} in which a high energy photon is produced at a large angle with respect to the beam direction; data in the energy range 189 - 208~GeV have been used. The selection procedures specific to this final state as well as to that in which a quark-antiquark pair is produced, rather than missing energy, are described in section~\ref{sec:exp_zg} below. 

In the final year of LEP running, data were taken over a range of energies from 205~to 208~GeV. The values of the centre-of-mass energy quoted in the descriptions below for that year correspond to the averages for the data samples collected.

\subsection{The $Z \gamma$ final state}
\label{sec:exp_zg}

The selection procedure for \Zg\ production in the kinematic region with greatest sensitivity to anomalous gauge couplings concentrated on a search for events with a  very energetic photon in the angular range $45\dgr <\theta_\gamma < 135\dgr$, where $\theta_\gamma$ is the polar angle of the photon with respect to the beam direction. This angular region is covered by DELPHI's barrel electromagnetic calorimeter, the High density Projection Chamber (HPC). The search was conducted in events with two final state topologies: \vvg\ and \qqg. 

\begin{figure}[h]
\centerline{\epsfig{file=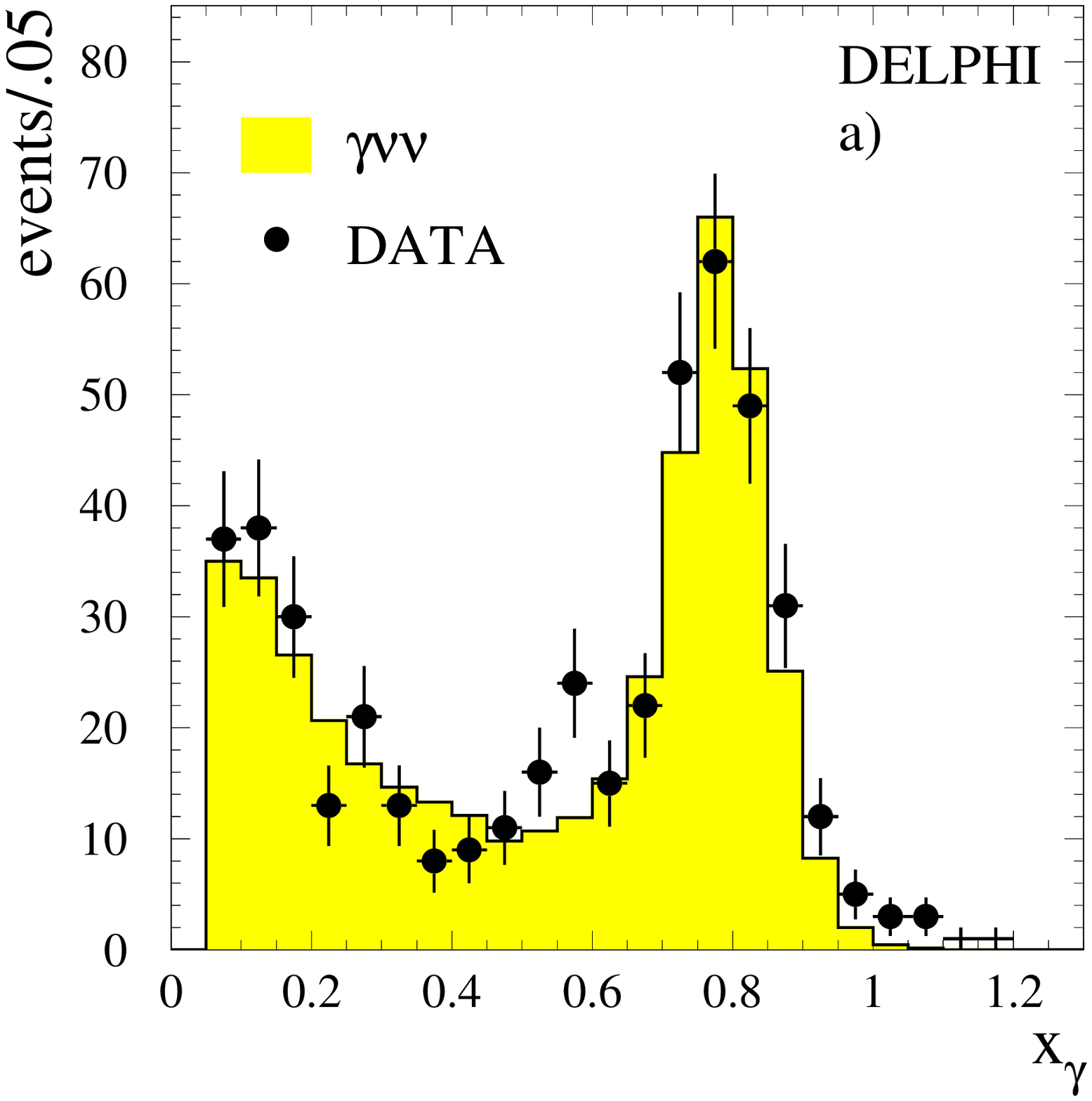,width=10cm}}
\vspace{0.5cm}
\centerline{\epsfig{file=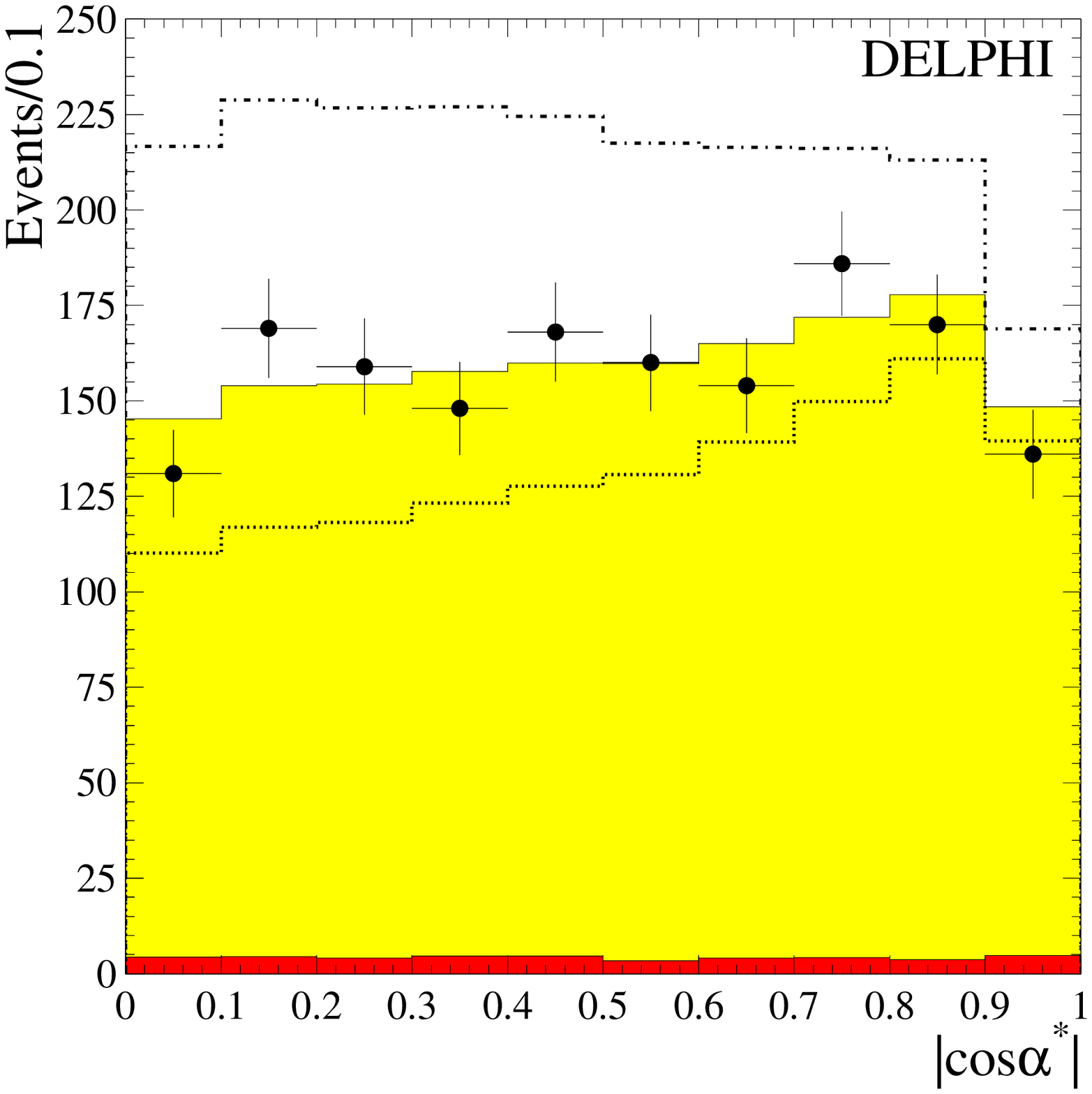,width=8cm} 
               \epsfig{file=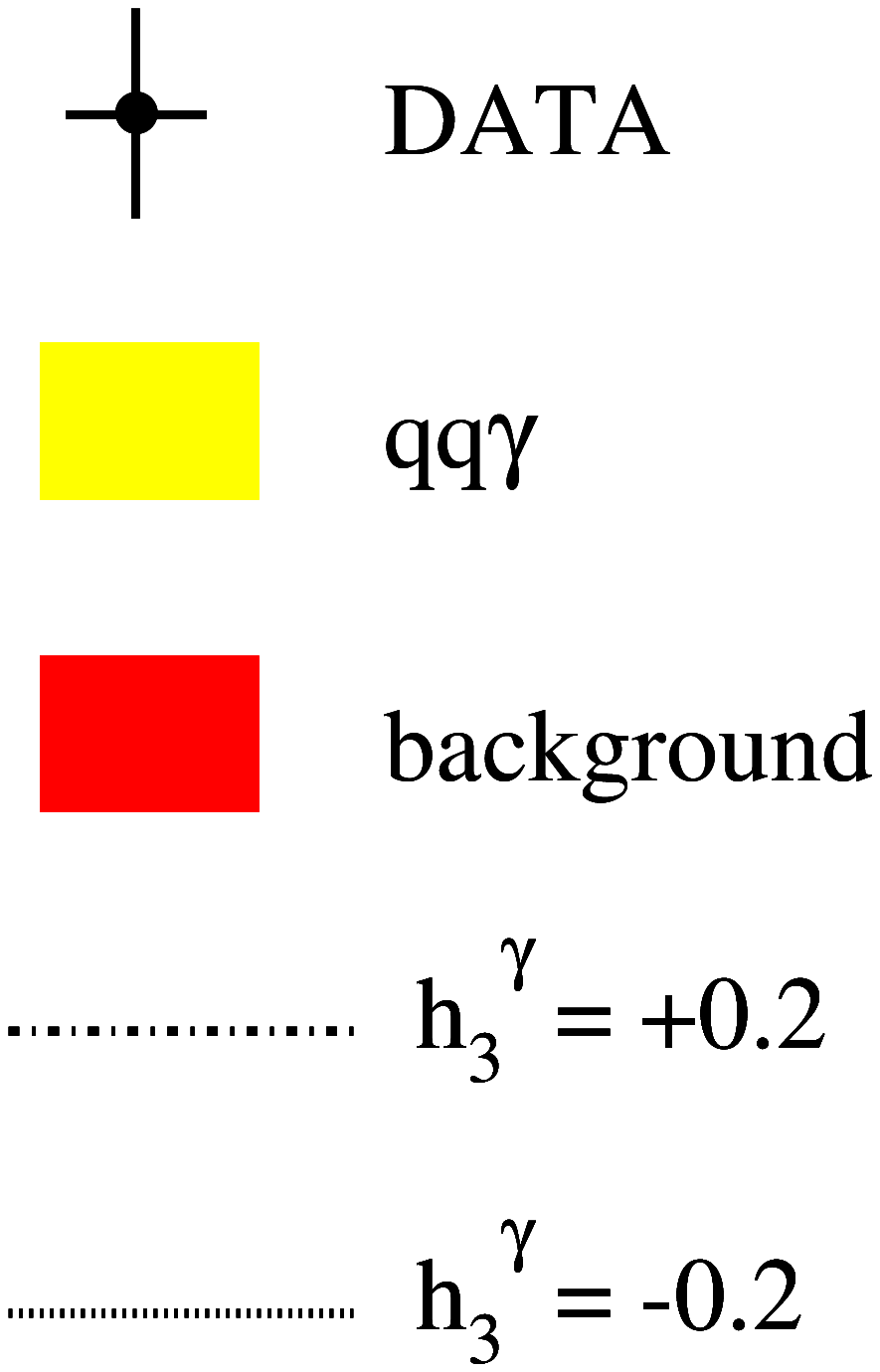,width=8.5cm}}
\vspace{-10.75cm}
\centerline{ \hspace{5.85cm} \Large{$\bullet$}}
\vspace{10.75cm}
\vspace{-11.25cm}
\centerline{ \hspace{6.3cm} \Large{$\bullet$}}
\vspace{11.25cm}
\vspace{-8.0cm}
\centerline{\hspace{-2.7cm}b)}
\vspace{8.0cm}
\vspace{-0.9cm}

\caption{a) Distribution of $x_\gamma = E_\gamma / E_{beam}$, the energy  of identified photons normalized to the beam energy in the data in the \vvg\ channel, summed over all energy points. The distribution is shown before imposing the experimental cut at $E_\gamma = 50$~GeV. The experimental data points are shown by dots and the shaded histogram shows the predictions of the Standard Model for signal plus background. (The background contribution is very small, and is not shown separately).    
b) Distribution of $|\cos\alpha^\star|$, where $\alpha^\star$ is the decay angle of the quark (or antiquark) in the $Z$ centre-of-mass frame with respect to the direction of the $Z$ in the overall centre of mass, for data selected in the \qqg\ channel. The experimental data points are shown by dots, the shaded histogram shows the predictions of the Standard Model for signal and background, and the outlined histograms the expectations for values of  $h_3^\gamma = \pm 0.2$. 
}
\label{fig:zg}
\end{figure}

The \vvg\ sample was selected from events with a detected final state containing only a single photon. Its energy, $E_\gamma$, was required to be greater than 50~GeV and only photons in the range covered by the HPC, $45\dgr <\theta_\gamma < 135\dgr$ were accepted.  No tracks or hits were allowed in the TPC. It was also required that no electromagnetic showers with energy exceeding defined background noise levels were present in the forward electromagnetic calorimeter and the luminosity monitor. Further showers in the HPC were accepted only if they were within 20\dgr\ of the first one, and such showers were then combined.  Cosmic ray events were suppressed by requiring that any signal in the hadronic calorimeter be in the same angular region as the signal in the electromagnetic calorimeter, and that the electromagnetic shower point towards the beam collision point within an angle of 15\dgr. The trigger efficiency was measured using Compton and Bhabha events. The expected numbers of events were calculated using the generators NUNUGPV, based on~\cite{nicro1}, and KORALZ~\cite{koralz}, interfaced to the full DELPHI simulation program. The results obtained applying these criteria are shown in table~\ref{table:vvg}. From the simulations, the efficiency for detecting \vvg\ events in the kinematic region considered here was shown to be independent of the centre-of-mass energy, with an average value of (50.7~$\pm$~2.0)\% for the data sample listed in table~\ref{table:vvg}. Contributions from background sources to this channel were estimated to be negligible. Full details of the analysis of this final state may be found in~\cite{gx}. The distribution of $x_\gamma$,  the energy of identified photons normalized to the beam energy ($x_\gamma = E_\gamma / E_{beam}$) before imposing the cut at $E_\gamma = 50$~GeV is shown for photons with $x_\gamma > 0.05$ in figure~\ref{fig:zg}a), which also shows the expectation of the Standard Model.

\begin{table}[htb]
\begin{center}
%\small{
\begin{tabular}{|c|c|c|c|}                                           \hline
$\sqrt{s}$ &Integrated            & Selected &Total predicted       \\ 
(GeV)      &luminosity (pb$^{-1}$)&  data    & events          \\ \hline
 188.6     &154.7                 &87        &89.2             \\ \hline
 191.6     &25.1                  &14        &13.1             \\ \hline
 195.5     &76.2                  &32        &37.5              \\ \hline  199.5     &83.1                  &45        &38.5              \\ \hline  201.6     &40.6                  &20        &18.2              \\ \hline  206.1     & 214.6                &98        &102.3     \\ \hline \hline
Total      & 594.3                &296       &298.8           \\ \hline         
\end{tabular}
%}
\caption{\vvg\ final state: Integrated luminosity and numbers of observed and expected events at each energy, $\sqrt{s}$.}
\label{table:vvg}
\end{center}
\end{table}

In the selection of events in the \qqg\ channel, the same requirements were imposed on the energy and polar angle of photon candidates as in the \vvg\ case, namely: $E_\gamma >$ 50~GeV and $45\dgr <\theta_{\gamma} < 135\dgr$. In addition, events were required to have  six or more charged  particle tracks, each with length greater than 20~cm, momentum greater than 200~MeV/$c$, polar angle between 10\dgr\ and 170\dgr, and transverse and longitudinal impact parameters at the interaction point of less than 4~cm. The total charged energy in the event was required to exceed 0.10$\sqrt{s}$ and the  effective energy of the collision~\cite{sprime}, excluding the detected photon, $\sqrt{s^\prime}$, was required to satisfy  $\sqrt{s^\prime} < $ 130~GeV. Jets were reconstructed using the LUCLUS~\cite{pythia} algorithm and, omitting the $\gamma$, the event was forced into a two-jet configuration. The identified photon was required to be isolated  from the nearest jet axis by at least 20\dgr. The efficiency, purity and the expected numbers of events from $q \bar{q} (\gamma)$ production were computed using events generated with PYTHIA~\cite{pythia}, relying on JETSET 7.4~\cite{pythia} for quark fragmentation, and interfaced to the full DELPHI simulation program. The results obtained applying this procedure are shown in table~\ref{table:qqg}. The efficiency for detecting \qqg\ events in the kinematic region considered here was found to be almost independent of the centre-of-mass energy for the data sample used, with an average value of (76.4~$\pm$~0.2)\%. The main background processes, contributing about 3\% of the sample, came from $q \bar{q}$ production with a photon from fragmentation of one of the quarks, and from $WW$ production.

\begin{table}[htb]
\begin{center}
%\small{
\begin{tabular}{|c|c|c|c|c|}                                           \hline
$\sqrt{s}$    &Integrated  & Selected &Total predicted & Expected      \\ 
(GeV)         & luminosity (pb$^{-1}$)
                                           &  data     &events              & background  \\ \hline
 188.6        &154.3      &454        &467.3        &  14.9  \\ \hline
 191.6        &25.4       &79         & 75.0        &  2.6       \\ \hline
 195.5        &77.1       &203        &214.1        &  5.8     \\ \hline      
 199.5        &84.2       &208        &225.5        &  5.9     \\ \hline      
 201.6        &40.6       &130        &104.5        &  2.8     \\ \hline 
 205.9        &218.8      &507        &515.1        & 13.9   \\ \hline \hline
Total         &600.4      &1581       &1601.5       &  45.9 \\ \hline     
\end{tabular}
\caption{\qqg\ final state: Integrated luminosity, numbers of observed and expected events and predicted background contribution at each energy, $\sqrt{s}$.}
\label{table:qqg}
\end{center}
\end{table}

Summing over all energy points, totals of 296 and 1581 events were observed in the \vvg\ and \qqg\ channels, respectively. These numbers may be compared with the totals expected from simulated production of these final states by Standard Model processes: 298.8 events in \vvg, and 1601.5 events in \qqg.

In the \vvg\ channel, values of the gauge boson coupling parameters were derived by comparing the observed number of events with that predicted from the total cross-section for this process, while in the \qqg\ channel a fit was performed to the distribution of $|\cos \alpha^\star|$, where  $\alpha^\star$ is the angle of the quark or antiquark from $Z$ decay in the $Z$ rest frame with respect to the direction of the $Z$ in the overall centre of mass. The value of $|\cos \alpha^\star|$ was estimated from the directions of the vectors in the laboratory frame $\bf{p}_\gamma$ and $\bf{p}_i$ ($i=1, 2$) of the reconstructed photon and jets, respectively, from the relation:

\begin{equation}
\cot \alpha^\star = \gamma \bigg ( \cot \alpha_1 - \frac{\beta}{\sin \alpha_1}
\bigg ) \, ,
\end{equation}
\begin{equation}
\mathrm{with} \ \ \beta = \frac{\sin(\alpha_1+\alpha_2)}{\sin\alpha_1 +
\sin\alpha_2}, \ \
\cos \alpha_i = - \frac{{\bf p}_\gamma \cdot {\bf p}_i}{|{\bf p}_\gamma| \cdot
|{\bf p}_i|} \ \ \mathrm{and} \ \
\gamma = \frac{1}{\sqrt{1-\beta^2}} \ .
\end{equation}

\noindent The distribution of $|\cos\alpha^\star|$ for the data selected in the \qqg\ channel is shown in figure~\ref{fig:zg}b) and compared with the predictions of the Standard Model and of a non-standard scenario with $h_3^\gamma = \pm0.2$. The predictions for non-zero neutral gauge boson couplings were made by reweighting the simulated samples produced according to the Standard Model with the calculations of Baur and Berger~\cite{baur}\footnote{The code used was modified by a factor $i$ according to the correction suggested by Gounaris {\it et al}~\cite{ifact}.}.

\subsection{The \ZZ\ final state}
\label{sec:exp_zz}

The study of the triple-gauge-boson vertex in \ZZ\ production used the samples of events selected in the \qqqq, \qqmm, \qqee, \qqvv, \mmvv\ and \eevv\ final states. The procedures used to extract the data have been described fully in~\cite{zz}; we give here a brief summary of the methods used in the selection of events in each of these final states, and provide a table of the total numbers of events observed and expected for production of each of them by Standard Model processes.

The \ZZ\  \ra\ \qqqq\ process represents 49\% of the \ZZ\ decay topologies and produces four or more jets in the final state. After a four-jet preselection, the \ZZ\ signal was identified within the large background from \WW\ and \qqg\  production by evaluating a  probability that each event came from \ZZ\ production, based on invariant-mass information, on the $b$-tag probability per jet and on topological information. 

The process \ee\ \ra\ \qqll\  has a branching ratio in \ZZ\ production of 4.7\% per lepton flavour. High efficiency and high purity were attained with a cut-based analysis using the clear experimental signature provided by the two leptons, which are typically well isolated from all other particles. The on-shell \ZZ\ sample was selected by applying simultaneous cuts on the masses of the $l^+l^-$ pair, on the remaining hadronic system and on their sum.

The decay mode \qqvv\ represents 28\% of the \ZZ\ final states. The signature of this decay mode is a pair of jets, acoplanar with respect to the beam axis, with visible and recoil masses compatible with the $Z$ mass. The most difficult backgrounds arise from single resonant $W e \nu_e$ production, from \WW\  production where one of the $W$ bosons decays into $\tau \nu_{\tau}$, and from $q\bar{q}$ events accompanied by energetic isolated photons escaping detection. The selection of events was made using a combined discriminant variable obtained with an Iterative Discriminant Analysis program (IDA)~\cite{ida}.
 
The final state \llvv\ has a branching ratio in \ZZ\ production of 1.3\% per charged lepton flavour. Events with $l \equiv \mu,e$ were selected with a sequential cut-based analysis. The on-shell \ZZ\ sample was selected from the events assigned to this final state by applying cuts on the masses of the $l^+l^-$ pair and of the system recoiling against it. The most significant background in the sample is from \WW\ production with both $W$s decaying leptonically.

In the estimation of the expected numbers of events in all the final states discussed above,  processes leading to a four-fermion final state were simulated with EXCALIBUR~\cite{excalibur}, with JETSET~7.4 used for quark fragmentation. Amongst the background processes leading to the 
final-state toplogies described above, GRC4F~\cite{grc4f} was used to simulate \Wev\ production, PYTHIA for \qqbrg, KORALZ for $\mu^+ \mu^- (\gamma)$ and $\tau^+ \tau^- (\gamma)$, BHWIDE~\cite{bhwide} for $e^+ e^- (\gamma)$, and TWOGAM~\cite{twogam} and BDK~\cite{bdk} for two-photon processes. 

The presence of anomalous neutral triple-gauge-boson couplings in the data samples described above was investigated by studying the distribution of the $Z$ production polar angle, $| \cos\theta_Z |$. For events in the \qqvv\ and \llvv\ final states, the $Z$ direction was taken to be the direction of the reconstructed di-jet or $l^+ l^-$ pair, respectively, while in the \qqll\ final state, the $Z$ direction was evaluated following a 4-constraint kinematic fit to the jet and lepton momenta, imposing four-momentum conservation. In the \qqqq\ final state,  the indistinguishability of the jets leads to three possible jet-jet pairs, each of which could come from \ZZ\ decay. A 5-constraint kinematic fit was performed on each of these combinations, imposing four-momentum conservation and equality of the masses of the two jet pairs. The fit with the minimum value of $\chi^2$ was retained and the value of $| \cos\theta_Z |$ evaluated from the fitted jet directions. 

\begin{figure}[ht]
\centerline{\epsfig{file=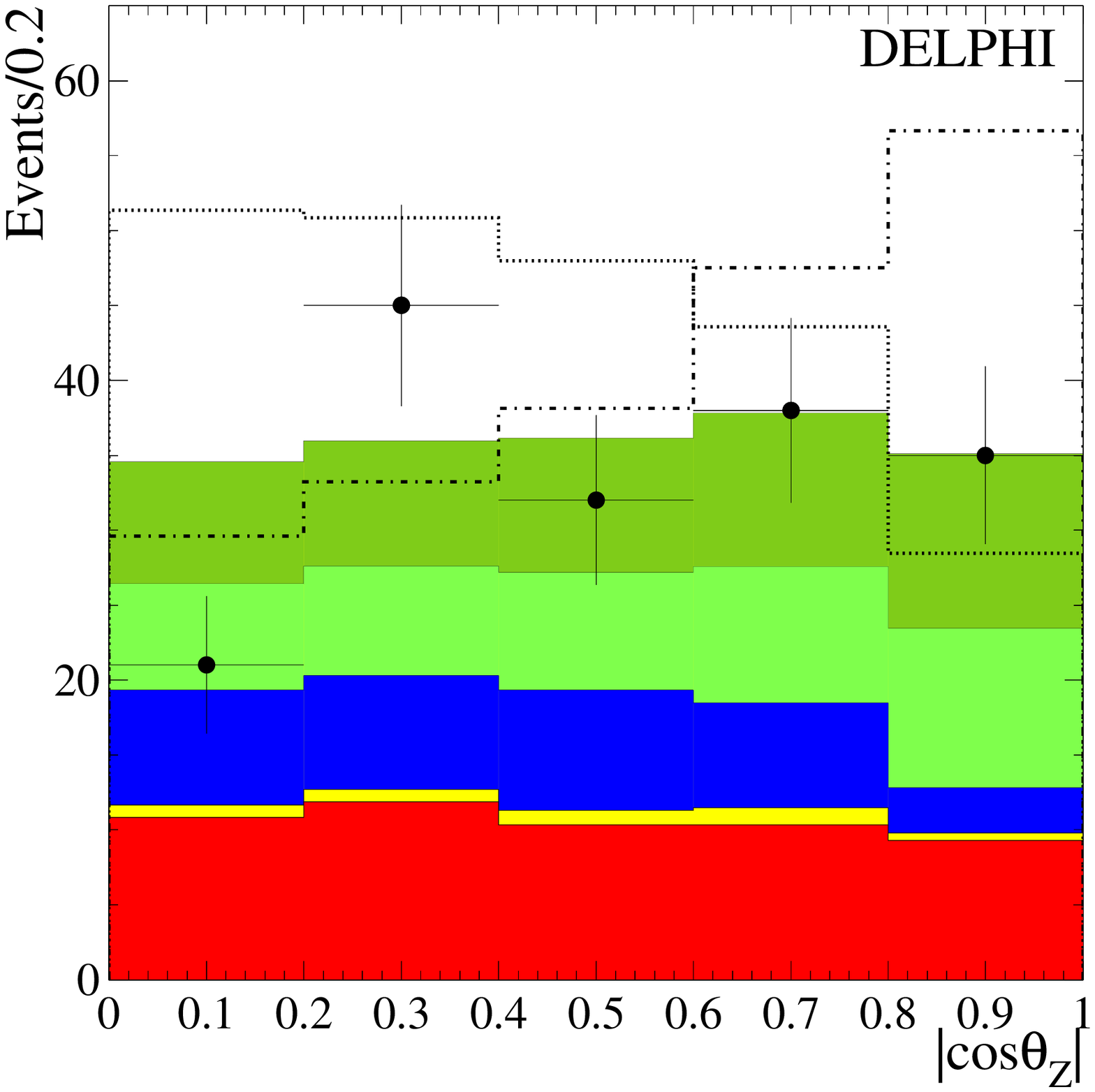,width=8cm}
            \epsfig{file=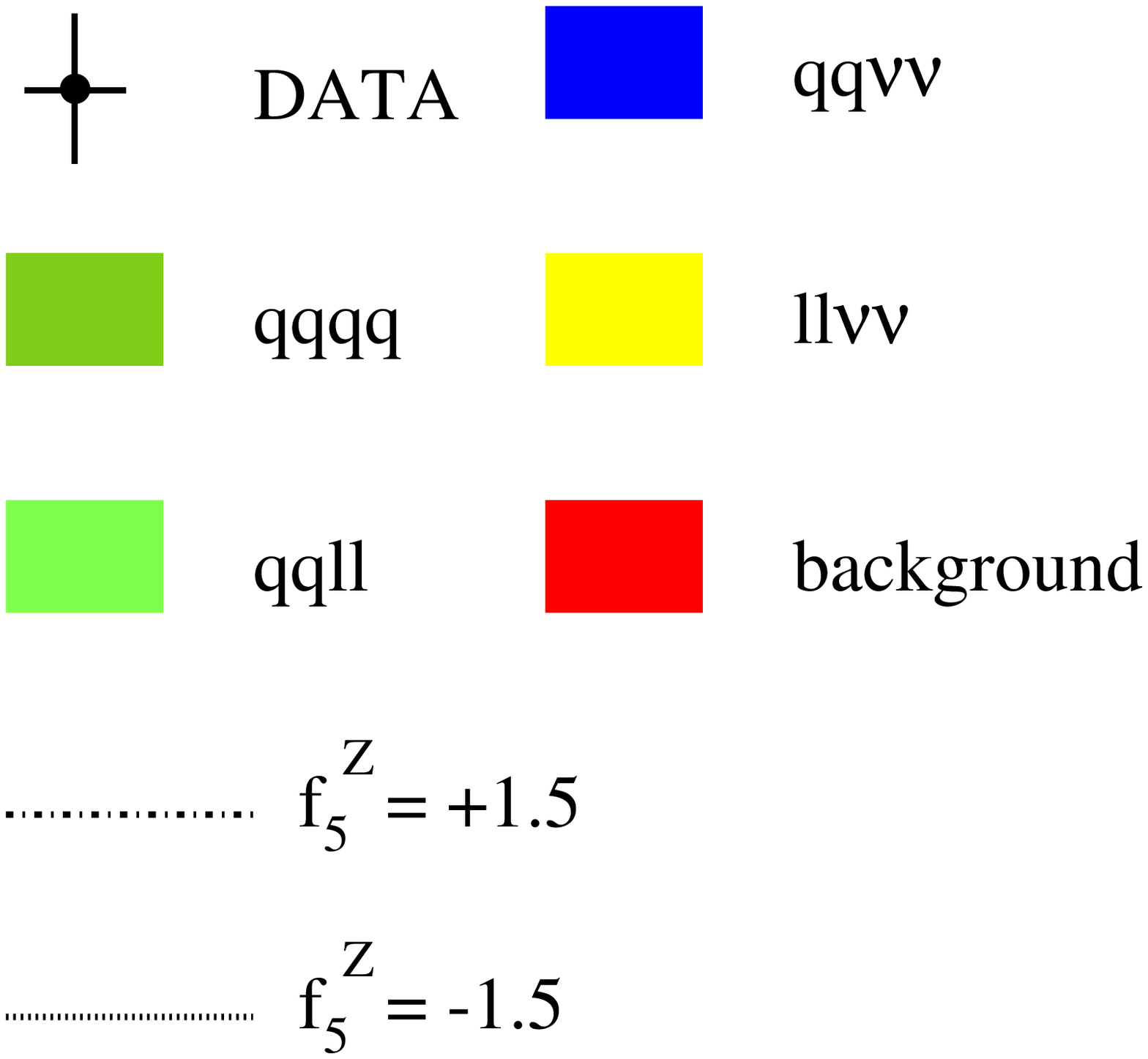,width=8cm} }
\vspace{0.9cm}
 \caption{Distribution of $| \cos\theta_Z |$, where $\theta_Z$ is the $Z$ production polar angle, for data selected in the \ZZ\ channels. The experimental data points are shown by dots, the shaded histograms show the predictions of the Standard Model for the signal and background components indicated in the legend, and the outlined histograms the expectations for values of $f_5^Z = \pm 1.5$.} 
\label{fig:zz}
\end{figure}

Figure~\ref{fig:zz} shows the distribution of $| \cos\theta_Z |$ for a high purity sample of \ZZ\ data, composed of the \qqll\ and \llvv\ samples defined above and, for illustrative purposes, samples of  \qqqq\ and \qqvv\  events defined by stringent cuts on the probabilistic variables used in these channels (\qqqq\ probability $> 0.55$, and \qqvv\ IDA variable $> 3$), so as to suppress the background levels present. (As described below, no cuts were imposed on these variables in the determination of coupling parameters). The figure also shows the Standard Model expectations and the distributions predicted for values of $f_5^Z = \pm 1.5$. The content of this sample is shown in table~\ref{table:zz}. The selection efficiencies for all of the channels analysed were found to have little energy variation over the range considered here, and the values quoted are averages for the whole experimental sample. They represent the fraction of events of the relevant four-fermion final state present in the selected sample, while the estimated backgrounds refer to contributions from other channels. 

\begin{table}[htb]
\begin{center}
%\small{
\begin{tabular}{|l|c|c|c|c|c|}                                                                                     \hline
Channel  &Integrated &Selected &Total predicted 
                                        & Expected & Selection \\ 
         & luminosity (pb$^{-1}$)
                     & data   &events & background & efficiency \\ \hline 
 \qqqq   &  665.1    &   76    &  69.4   &  22.1   & 0.18       \\ \hline
 \qqmm   &  665.3    &   21    &  22.0   &   1.1   &   0.86     \\ \hline
 \qqee   &  665.3    &   19    &  23.7   &   2.6   &    0.73    \\ \hline 
 \qqvv   &  639.0    &   45   &   55.5   &  22.3  &     0.21   \\ \hline
  \llvv   &  665.3    &  10    &   8.9   &   4.7   &     0.30   \\ \hline                                                                   \hline
Total    &   --      &  171   &  179.5   &  52.8    &    --      \\ \hline \end{tabular}
%}
\caption{\ZZ\ production: Integrated luminosity, numbers of observed and expected events and predicted background contribution for each topological final state, summed over all energies. The last column shows the energy-averaged selection efficiency for each final state.
}
\label{table:zz}
\end{center}
\end{table}

In the determination of the coupling parameters, extended maximum-likelihood fits were made to the distribution of $| \cos\theta_Z |$ for data from the channels selected with cut-based analyses (\qqll\ and \llvv), while for the channels selected using probabilistic methods (\qqqq\ and \qqvv), a simultaneous fit was made to the distributions of  $| \cos\theta_Z |$ and of the discriminant variable (the \ZZ\ probability for \qqqq\ and the IDA output variable for \qqvv), without applying any cuts on the values of these variables.

The predictions for non-zero neutral gauge boson couplings were made by reweighting the simulated samples produced according to the Standard Model with the calculations of the DELTGC~\cite{deltgc} event generator, which adds the amplitude from hypothesized neutral triple-gauge-boson vertices to all the other amplitudes contributing to the production of any four-fermion final state.

\subsection {The \Zg*\ final state}
\label{sec:exp_zgst}

In a separate publication~\cite{zgst}, DELPHI has reported on a study of \Zg*\ production in LEP2 data, and in particular on a comparison of the observed cross-section with Standard Model predictions, using data from a variety of four-fermion final state topologies involving both hadronic and leptonic $Z$ decay modes. In the present analysis, we interpret data in the \qqmm\ and \qqee\ final states in terms of possible anomalous triple-gauge-boson interactions. These two channels are chosen because the two final state leptons are typically well isolated from all other particles, allowing such events to be selected with high efficiency over the whole region of $\gamma^*$ mass. Events with either the $l^+l^-$ or the $q \bar{q}$ invariant mass in the vicinity of the $Z$ mass and the other invariant mass not in the $Z$ region were then used in the estimation of possible anomalous gauge coupling parameters. Full details of the selection procedure are given in~\cite{zgst}; a summary of the main features follows.

\begin{figure}[ht]
\centerline{\epsfig{file=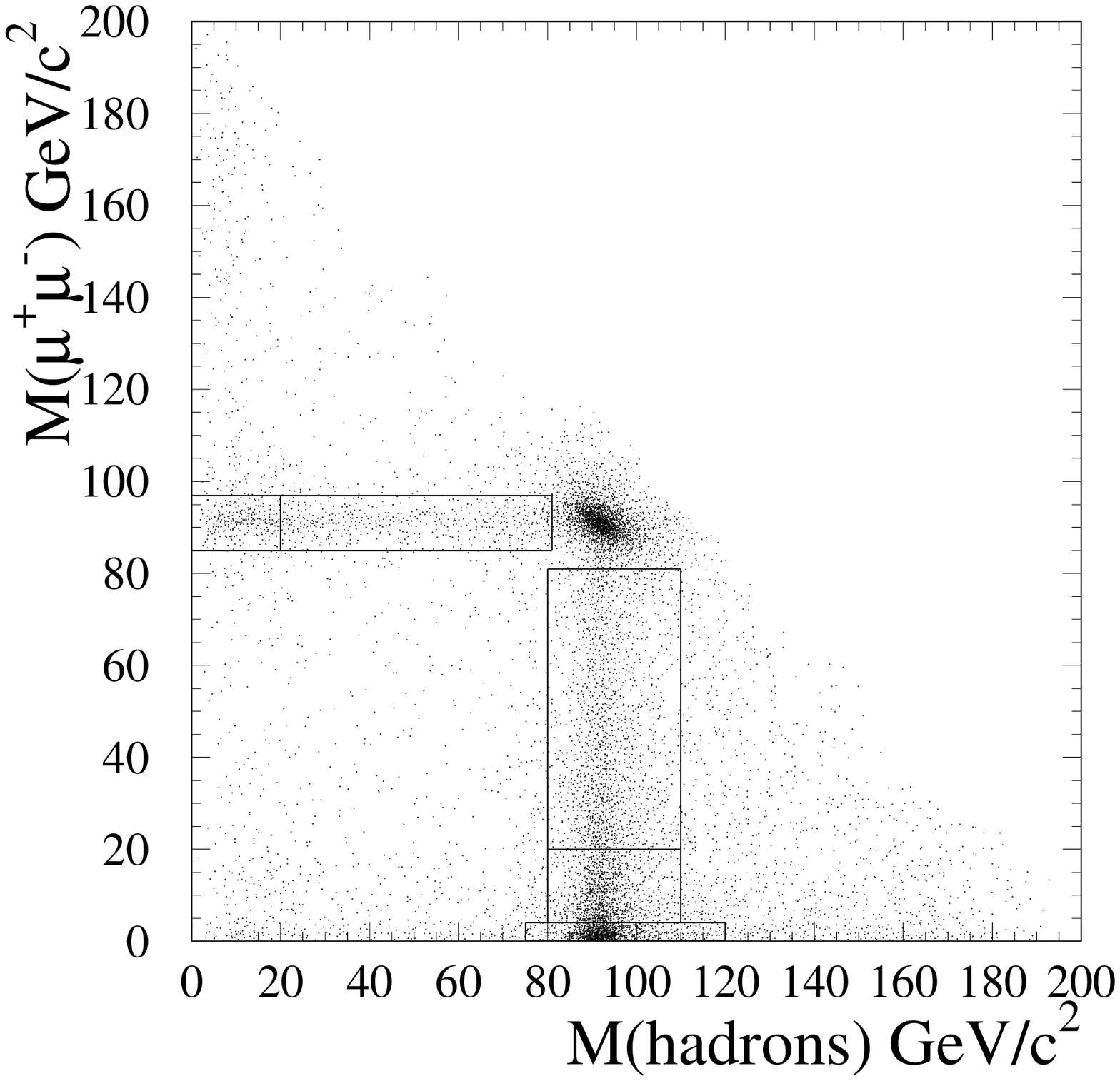,width=8cm}
                \epsfig{file=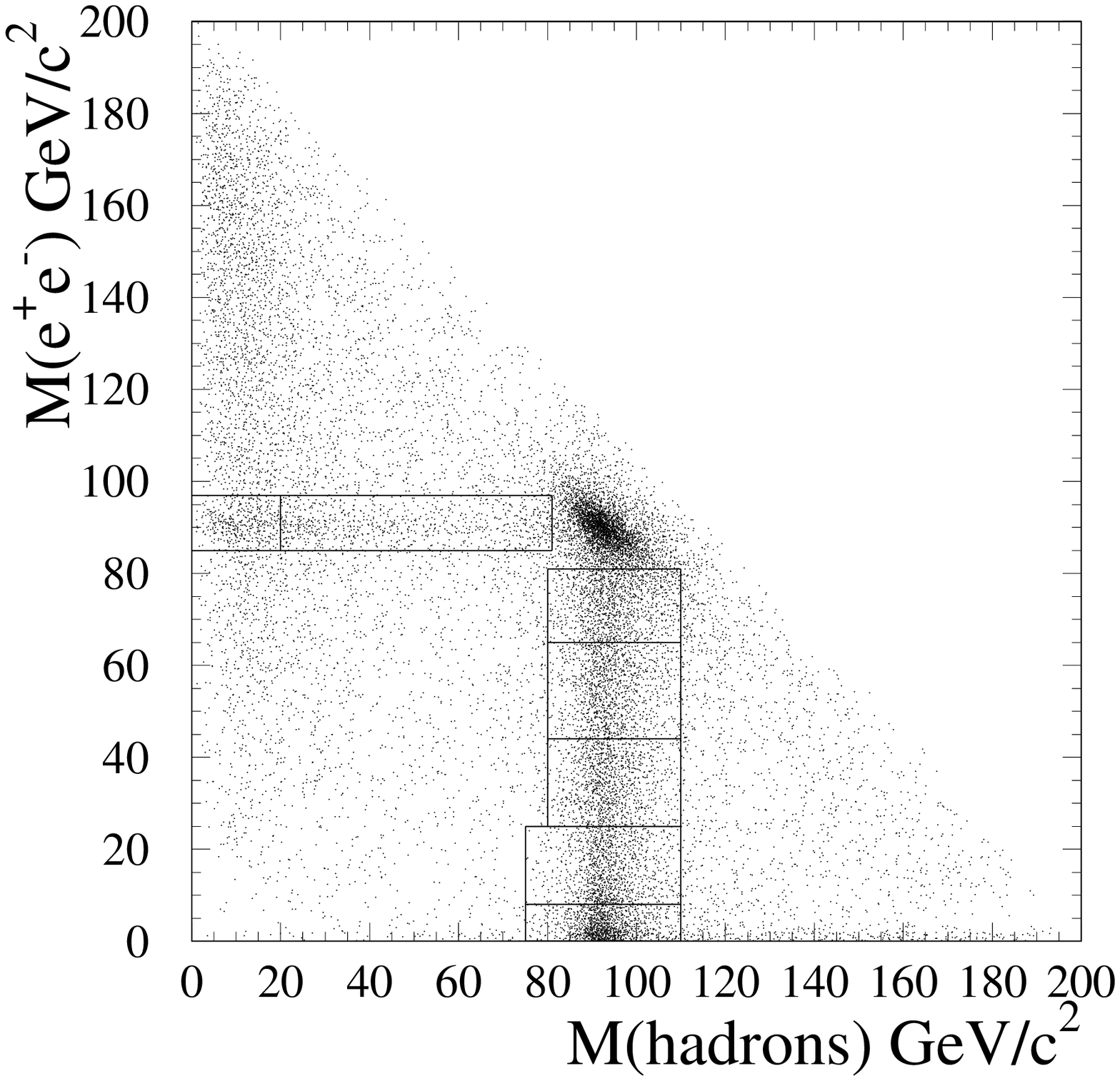,width=8cm}}
\vspace{1cm}
\centerline{\epsfig{file=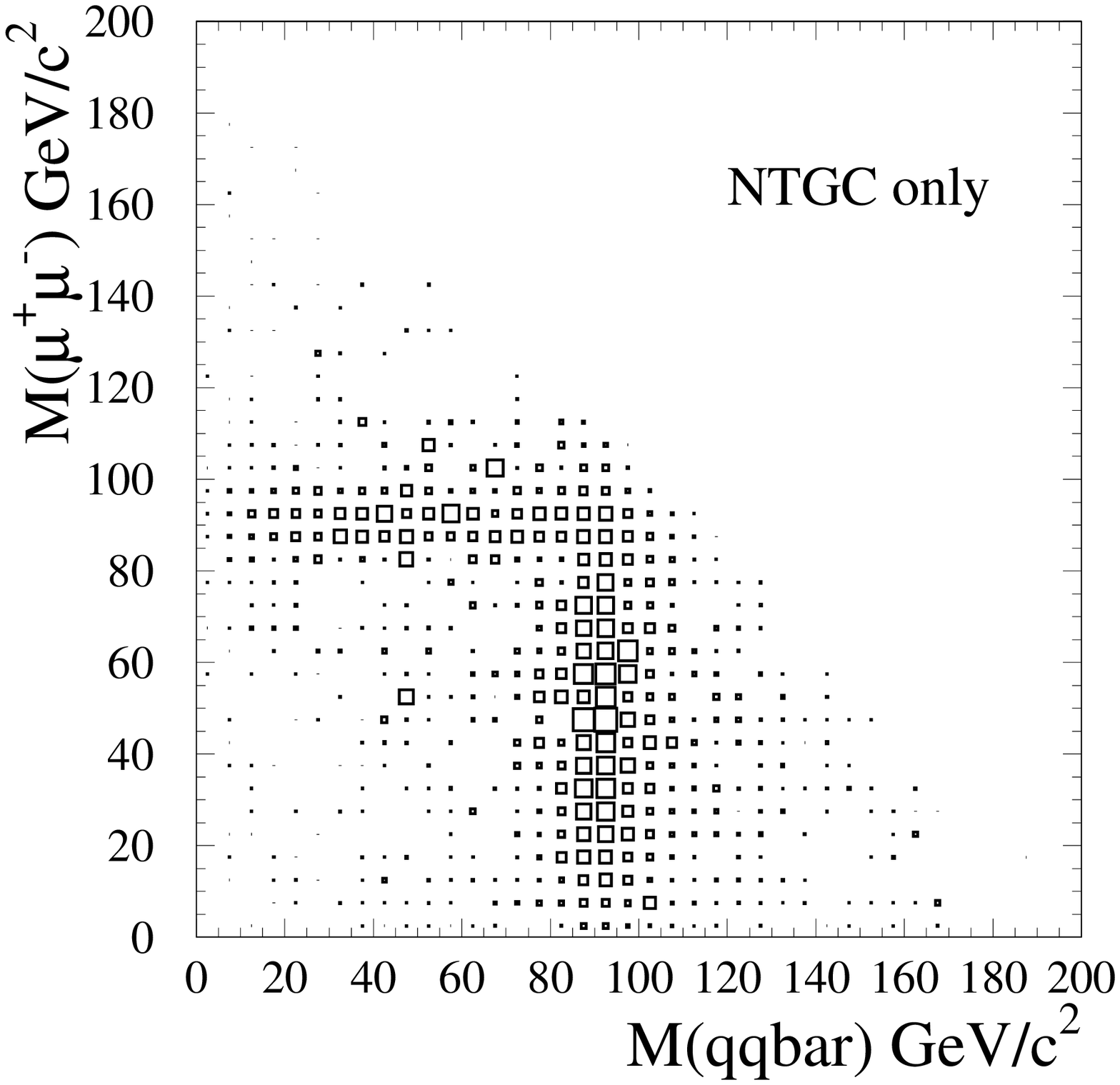,width=8cm}
            \epsfig{file=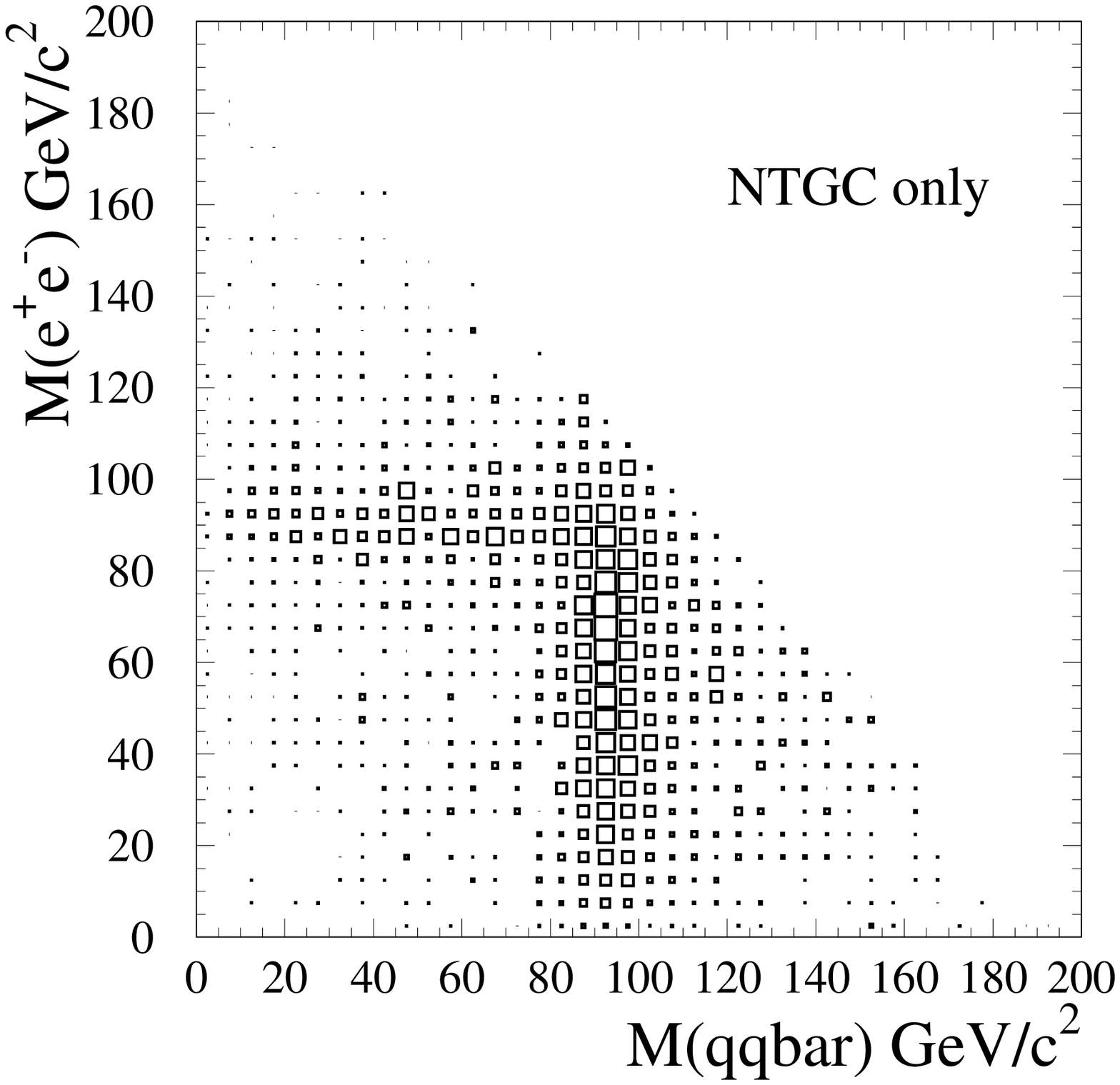,width=8cm}}
\vspace{-17.5cm}
\centerline{\hspace{-7.0cm} \bf{a)} \hspace{7.1cm} \bf{b)}}
\vspace{17.5cm}
\vspace{-8.8cm}
\centerline{\hspace{-7.0cm} \bf{c)} \hspace{7.1cm} \bf{d)}}
\vspace{8.8cm}
\caption{ 
For the \qqll\ final state: predicted Standard Model distributions of events a) in the ($M_{hadrons}$, $M_{\mu^+ \mu^-}$) plane, and b) in the ($M_{hadrons}$, $M_{e^+ e^-}$) plane, showing the bins used in the fits to the coupling parameters. The sum of all the bins defines the \Zgst\ sample.  c) Expected distribution in the ($M_{q \bar q}$, $M_{\mu^+ \mu^-}$) plane, and~d) in the ($M_{q \bar q}$, $M_{e^+ e^-}$) plane, of the difference between the predictions of the Standard Model plus an anomalous contribution, $\tilde{\ell}_4^{\ZZg} m_Z^4= 3.4$, and the Standard Model only. (The parameter $\tilde{\ell}_4^{\ZZg}$ is defined in table~\ref{table:parameters}). Plots~a) and~b) and, separately,~c) and~d) were computed with the same assumed luminosities.
}
\label{fig:zgstbins}
\end{figure}

Events containing  total charged hadronic energy above 0.3$\sqrt{s}$ and at least two lepton candidates of the same flavour and opposite charge were selected. All particles except the lepton candidates were clustered into jets and a kinematic fit requiring four-momentum conservation was applied. At least one of the two lepton candidates was required to satisfy strong lepton identification criteria, while softer requirements were specified for the second. In order to increase the purity of the selection, further cuts were made in two discriminating variables:  $P_t^{min}$, the lesser of the transverse momenta of the lepton candidates with respect to their nearest jet, and the $\chi^2$ per degree of freedom of the kinematic fit. This procedure selected a total of 170 events in the combined \qqmm\ and \qqee\ channels. The \Zgst\ sample was then defined within the selected \qqll\ data by requiring the mass of one and only one $f \bar{f}$ pair to be in the $Z$ region. This was effected by imposing mass cuts in the $(M_{hadrons}, M_{\mu^+ \mu^-})$ and $(M_{hadrons}, M_{e^+ e^-})$ planes, where $M_{hadrons}$ represents the mass of the $q \bar{q}$ pair estimated from the reconstructed hadronic data; these cuts are defined in figures~\ref{fig:zgstbins}a) and~b) for the \qqmm\ and \qqee\ samples, respectively. Table~\ref{table:zgst} summarizes the selection procedures outlined above, showing, for the sum of data over all energy points, the total integrated luminosity, the numbers of observed and predicted events in the \Zgst\ region, defined as described above, and the estimated selection efficiency (defined as for the \ZZ\ sample described in section~\ref{sec:exp_zz} above) for each topological final state. The selection efficiencies for the \qqmm\ and \qqee\ samples analyzed here were shown to be independent of the centre-of-mass energy, with average values of  (44.1~$\pm$~0.2)\% and  (28.7~$\pm$~0.1)\%, respectively. The backgrounds in the selected samples are small, coming mainly from \qqtt, \WW\ and, in the case of  \qqee, from \qqbrg\ production. In the estimation of backgrounds and selection efficiency, the simulation of processes leading to four-fermion final states was done with WPHACT~\cite{wphact}, using the JETSET  model for quark hadronization, while the \qqbrg\ final state was simulated with the KK2f~\cite{kk2f} model. Both of these programs were interfaced to the DELPHI simulation program. 

\begin{table}[htb]
\begin{center}
%\small{
\begin{tabular}{|l|c|c|c|c|c|}                                                                                     \hline
Channel&Integrated &Selected &Total predicted & Expected & Selection \\ 
       & luminosity (pb$^{-1}$)
                   &  data   &events          &  background & efficiency                                                                  \\ \hline 
 \qqmm &   666.7    &  35    & 36.7           & 3.4 &  0.44    \\ \hline
 \qqee &   666.7    &  39    & 36.3           & 6.0 &    0.29    \\ \hline \end{tabular}
%}
\caption{\Zgst\ production: Integrated luminosity, numbers of observed and expected events and predicted background contribution for each topological final state, summed over all energies. The last column shows the energy-averaged selection efficiency for each final state.
}
\label{table:zgst}
\end{center}
\end{table}

\begin{figure}[ht]
\centerline{\epsfig{file=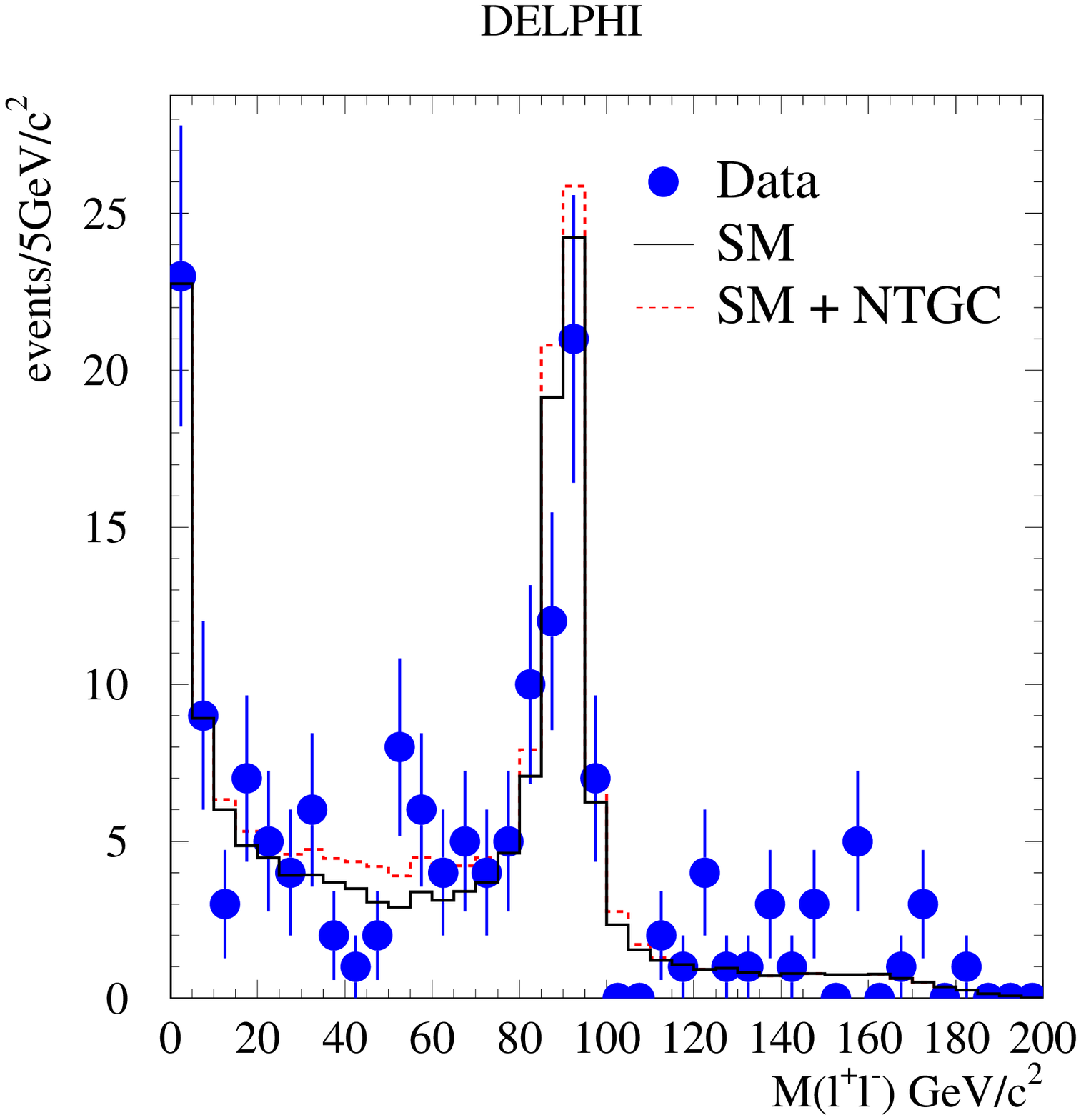,width=8cm}
            \epsfig{file=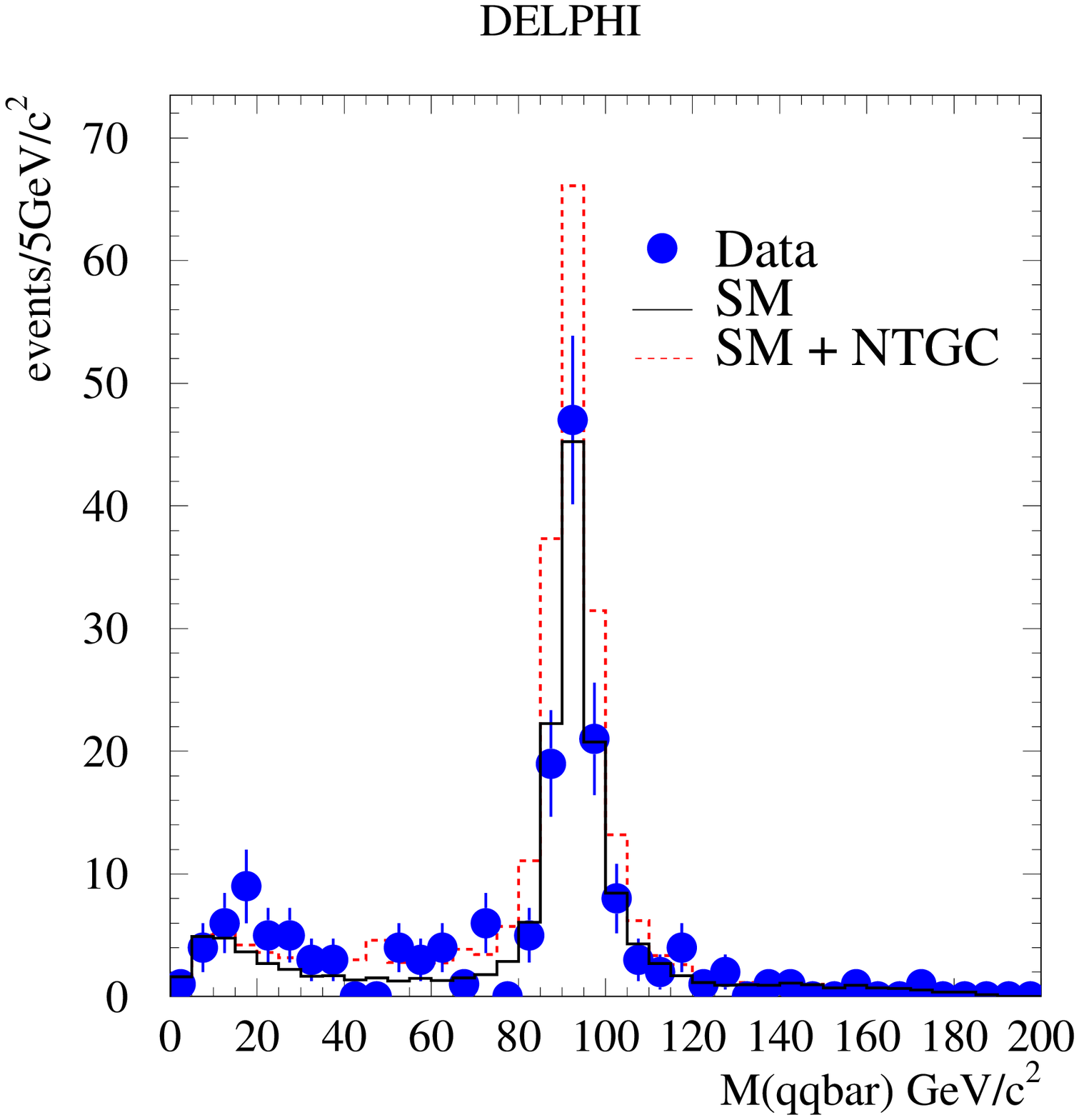,width=8cm}}
\vspace{-8.4cm}
\centerline{\hspace{-7.0cm} \bf{a)} \hspace{7.1cm} \bf{b)}}
\vspace{8.4cm}
\caption{ 
a) Distribution of $M_{l^+ l^-}$ ($l \equiv e,\mu$), and b)~of $M_{q \bar{q}}$, for data selected in the \qqmm\ and \qqee\ channels. The experimental data points are shown by dots, the full histograms show the predictions of the Standard Model for signal and background, and the dotted histograms the expectations when an anomalous contribution, $\tilde{\ell}_4^{\ZZg} m_Z^4= 3.4$, is present. 
}
\label{fig:zgst}
\end{figure}

Some aspects of the phenomenology of \qqll\ production in the context of possible neutral triple-gauge-boson couplings, and of the data selected in the \qqmm\ and \qqee\ channels, are demonstrated in figures~\ref{fig:zgstbins} and~\ref{fig:zgst}. Figures~\ref{fig:zgstbins}a) and~b) show the distributions in the $(M_{hadrons}, M_{l^+ l^-})$ planes predicted by the Standard Model for the \qqmm\ and \qqee\ final states, respectively. These differ considerably, due to the presence of additional diagrams contributing to \qqee\ production, in particular those corresponding to the production of $Z$\ee\ and $\gamma^*$\ee\ by $t$-channel processes. These effects have been discussed fully in~\cite{zgst}. The effect of an anomalous triple-gauge-boson coupling in these channels is illustrated in figures~\ref{fig:zgstbins}c) and~d), which show, respectively, the difference between the expected distributions on the ($M_{q \bar{q}}$, $M_{\mu^+ \mu^-}$) and ($M_{q \bar{q}}$, $M_{e^+ e^-}$) planes when a non-zero contribution from the $d=8$ operator ${\tilde{\cal O}}_4^{\ZZg}$ (defined in table~\ref{table:parameters}) is included, and when only the Standard Model amplitudes are used. Again, some differences between the predictions for the \qqmm\ and \qqee\ final states are observed; these are due to the presence of additional diagrams in the \qqee\ amplitude, in this case the $V^0 V^0$ fusion diagram leading to $Z$\ee\ production, shown in figure~\ref{fig:diagrams}d). The overall effect is a negative interference between $s$- and $t$-channel amplitudes: for the example shown, the predicted content of figure~\ref{fig:zgstbins}d) (\qqee) is $\sim 40\%$ of that of the \qqmm\ prediction. 

Data selected over the whole region of the \qqll\ phase space are presented in figures~\ref{fig:zgst}a) and~b) in the form of the distributions of $M_{l^+ l^-}$ ($l \equiv \mu, e$) and $M_{q \bar{q}}$. These plots also show the expectations of the Standard Model and of a model in which an anomalous contribution $\tilde{\ell}_4^{\ZZg} m_Z^4= 3.4$ from the operator ${\tilde{\cal O}}_4^{\ZZg}$ is present.

In the determination of the coupling parameters, the regions in the plane of the masses of the two fermion-antifermion pairs defining the \Zgst\ samples in the \qqmm\ and \qqee\ final states, shown in figures~\ref{fig:zgstbins}a) and~b), respectively, were divided into a small number of bins of unequal size, but containing roughly equal numbers of events predicted by the Standard Model. Different bin definitions were made for the two channels; the bins are also defined in the figures, and they  correspond to those used by DELPHI in~\cite{zgst} in the determination of the \Zgst\ cross-section. In~\cite{zgst}, each of the mass bins defined for the \qqee\ event sample was further divided into two angular regions, ($40^\circ < \theta_{l^+l^-} <140^\circ$) and ($\theta_{l^+l^-} < 40^\circ$ or $\theta_{l^+l^-} > 140^\circ$), where $\theta_{l^+l^-}$ is the polar angle of the final state $e^+e^-$ system with respect to the beam direction. These angular regions correspond to DELPHI's barrel and endcap regions, respectively. In the present analysis, we have extended this division to apply to muon as well as electron pairs in the \qqll\ final states. Binned likelihood fits to the couplings were then made with the bins in $(M_{q \bar{q}}, M_{l^+ l^-})$ and $\theta_{l^+ l^-}$ thus defined.  As in the case of the \ZZ\ final state previously described, the predictions for non-zero neutral gauge boson couplings in the \Zgst\ data were made by reweighting the simulated samples produced according to the Standard Model with the calculations of DELTGC.

\section{Results}
\label{sec:results}

In this section the results of our study are presented, expressed in terms of the parameters listed in table~\ref{table:parameters} describing the neutral triple-gauge-boson effective Lagrangian. In summary, these parameters represent:
\begin{itemize}
\item[a)] the coefficients of the lowest dimension operators contributing to production either of the \ZZ\ and \Zgst\ final states, or to production of the \Zg, \Zgst\ and \ZZ\ final states; in the on-shell \ZZ\ or \Zg\ limit each of these parameters becomes equal to one of the on-shell coefficients $f^V_i$ or $h^V_i$; 
\item[b)] the coefficients of the lowest dimension operators affecting only the $V^0 Z \gamma^*$ vertex; 
\item[c)] the coefficients of the $SU(2) \times U(1)$-conserving operators describing the \VVV\ vertex in i)~the Gounaris-Layssac-Renard and ii)~the Alcaraz formulations. 
\end{itemize}
\noindent The labels a), b), c) above refer to table~\ref{table:parameters}.

Limits on the parameters at the 95\% confidence level are given in table~\ref{table:results} and the corresponding likelihood curves are shown in figures~\ref{fig:fi}-\ref{fig:alcaraz}. In all cases, the values quoted are derived from one-parameter fits to the data in the \Zg, \ZZ\ and \Zgst\ channels described in sections~\ref{sec:exp_zg}, \ref{sec:exp_zz} and~\ref{sec:exp_zgst} above, summing the distributions from different channels where appropriate. In each fit, the values of the other parameters were set to zero, their Standard Model value. The results shown include contributions from both statistical and systematic effects. 

For reference, we summarize here the composition of the likelihood function from each of the final states used in the analysis, described in more detail in the sections above: In the \Zg\ \ra\ \vvg\ channel, the number of events with a high energy photon emitted at large polar angle was used in the fit, while in the \Zg \ra\ \qqg\ channel the fit was performed to the distribution of the decay angle of the $Z$ in its rest frame. In the channels \ZZ\ \ra\ \qqll\ and \ZZ\ \ra\ \llvv\ the distribution of the $Z$ production angle was fitted; in \ZZ\ \ra\ \qqqq\ and \ZZ\ \ra\ \qqvv\ simultaneous fits were made to the $Z$ production angle and, respectively, to the event probability or discriminant variable distributions. In the \Zgst\ channels studied (\Zgst\ \ra\ \qqmm\ and \Zgst\ \ra\ \qqee) the likelihood was evaluated in bins of $q \bar q$ or $l^+ l^-$ mass and of the polar angle of the detected  $l^+ l^-$ system. 

It may be noted that, in the models conventionally used to describe anomalous gauge-boson couplings, including the one used in this paper, all observables have a quadratic dependence on the fitted parameters. This effect, which has been previously noted (see, for example,~\cite{opal1}), leads to log-likelihood distributions which can have double mimima, asymmetries, and a broadening compared with that expected in the Gaussian case. Such features are indeed seen in several of the plots in figures~\ref{fig:fi}-\ref{fig:alcaraz}. The confidence limits reported in table~\ref{table:results} must therefore be interpreted with this effect in mind.

\begin{table}
\begin{center}
%\small{
\begin{tabular}{|c|c|l|c|}
\hline 
Parameter                            &Channels           & 95\%  Confidence  &  Related on-shell     \\
                                           &  used                &\ \ \ \ \  interval        &  coefficient             \\ \hline \hline
\mco{4}{|l|}{a)}                                                                               \\ \hline
$\tilde{\ell}_1^{\ZZZ} m_Z^2$        &  \ZZ\ \Zgst         
 &$[-0.40,+0.42]$      & $f_4^Z$      \\
$\ell_1^{\ZZZ} m_Z^2$                &  \ZZ\ \Zgst             
 & $[-0.38,+0.62]$     & $f_5^Z$      \\ 
$\tilde{\ell}_3^{\ZZg} m_Z^2$        &  \ZZ\ \Zgst         
 &$[-0.23,+0.25]$      & $f_4^\gamma$ \\
$\ell_2^{\ZZg} m_Z^2 $               &  \ZZ\ \Zgst             
 & $[-0.52,+0.48]$     & $f_5^\gamma$  \\ \hline
$\tilde{\ell}_1^{\ZZg} m_Z^2$        &  \Zg\ \Zgst\ \ZZ         
 & $[-0.23,+0.23]$     & $h_1^Z$      \\
$\ell_1^{\ZZg} m_Z^2 $               &  \Zg\ \Zgst\ \ZZ             
 & $[-0.30,+0.16]$     & $h_3^Z$       \\ 
$\tilde{\ell}_1^{\Zgg} m_Z^2$        &  \Zg\ \Zgst\ \ZZ         
 & $[-0.14,+0.14]$     & $h_1^\gamma$ \\
 $\ell_1^{\Zgg } m_Z^2$               &  \Zg\ \Zgst\ \ZZ     
 & $[-0.049,+0.044]$   & $h_3^\gamma$ \\ \hline \hline
\mco{4}{|l|}{b)}                      \\ \hline
$\tilde{\ell}_4^{\ZZg}  m_Z^4$       & \ZZ\ \Zgst\                    
 & $[-1.67,+1.92]$    &   --          \\  
$\ell_2^{\Zgg} m_Z^4$                & \ZZ\ \Zgst\                    
  & $[-0.49,+0.61]$     &  --      \\ \hline \hline
\mco{4}{|l|}{c)\ i)}                                                                                  \\ \hline
$- \cot \theta_W m_Z^2 \frac{v^2}{4} \tilde{\ell}_{SU(2) \times U(1)}$
                                     & \Zg\ \Zgst\ \ZZ         
 & $[-0.13,+0.13]$    & $h_1^\gamma$ \\
$- \cot \theta_W m_Z^2 \frac{v^2}{4} \ell_{SU(2) \times U(1)}$
                                     & \Zg\ \Zgst\ \ZZ        
 & $[-0.045,+0.047]$   & $h_3^\gamma$ \\ \hline
\mco{4}{|l|}{\ \ \ ii)}                                                                   \\ \hline
$- \cot \theta_W m_Z^2 \frac{v^2}{4} \tilde{\ell}_8^A$
                                     & \Zg\ \Zgst\ \ZZ         
 & $[-0.14,+0.14]$   & $h_1^\gamma$   \\
$- \cot \theta_W m_Z^2 \frac{v^2}{4} \ell_8^A$
                                     & \Zg\ \Zgst\ \ZZ       
 & $[-0.049,+0.045]$   & $h_3^\gamma$   \\  
$- \cot \theta_W m_Z^2 \frac{v^2}{4} \tilde{\ell}_8^B$ 
                                     & \Zg\ \Zgst\ \ZZ       
 & $[-0.23,+0.24]$   & $h_1^Z$ \\
$- \cot \theta_W m_Z^2 \frac{v^2}{4} \ell_8^B$
                                     & \Zg\ \Zgst\ \ZZ     
 & $[-0.30,+0.18]$ & $h_3^Z$ \\ \hline
\end{tabular}
%}
\caption{
Results of the study of neutral gauge couplings. For each of the parameters listed in table~\ref{table:parameters}, the table shows the experimental channels used and the 95\% confidence limits obtained. The right-most column indicates the parameter which, in the on-shell limit, is equal to the parameter determined. In the determination of any one coupling, the values of all the others were held at their Standard Model values. The limits shown include both statistical and systematic effects: a)~Coefficients of lowest dimension operators contributing either to \ZZ\ and \Zgst\ production or to \Zg, \Zgst\ and \ZZ\ production; b)~Coefficients of lowest dimension operators affecting only the $V^0 Z \gamma^*$ vertices; c)~Coefficients of $SU(2) \times U(1)$-conserving operators according to i) the Gounaris-Layssac-Renard constraints and ii) the Alcaraz constraints (see text, section~\ref{sec:intro_phen}).
 }
\label{table:results}
\end{center}
\end{table}

\subsection{Systematic errors}
\label{sec:systematics}

In the determination of the confidence limits shown in table~\ref{table:results} and the likelihood curves of figures~\ref{fig:fi}-\ref{fig:alcaraz}, several sources of systematic error were considered for each of the final states included in the study. These are described below.

In the \vvg\ and \qqg\ channels contributing to \Zg\ production, uncertainties of $\pm 1\%$ were assumed in the values assumed for the Standard Model production cross-sections~\cite{nicro1,koralz}, and an experimental uncertainty of $\pm 1\%$ was assumed for the energy calibration of the electromagnetic calorimeter.  The effect of an uncertainty of $\pm 1\%$ in the luminosity measurement was also computed, while the uncertainties in the calculations arising from the finite simulated statistics in signal and background channels and from the uncertainty in the knowledge of the background cross-section were found to be negligible in both channels. In the \vvg\ channel, the error due to the uncertainty of $\pm 3\%$ in the trigger efficiency was included. In the \qqg\ channel, the uncertainty in the use of PYTHIA as the hadronization model was taken into account by comparing events simulated with PYTHIA and HERWIG~\cite{herwig}; this gave rise to an estimated systematic error on the selection efficiency of $\pm 1.7\%$ from this source. In the combination of data at different energies, all the above effects were considered as correlated. The resulting overall systematic error in the coupling parameters was found to be of the order of  30\% of the statistical errors in the case of $h_1^Z$ and $h_3^Z$, about 50\% of the statistical error for $h_1^\gamma$, and of the same order as the statistical error for $h_3^\gamma$. In combination with \Zgst\ and \ZZ\ data to produce the limits on the parameters $\tilde{\ell}_1^{\ZZg} m_Z^2$, $\ell_1^{\ZZg} m_Z^2 $, $\tilde{\ell}_1^{\Zgg} m_Z^2$ and  $\ell_1^{\Zgg } m_Z^2$ shown in table~\ref{table:results}a) the \Zg\ data dominate (see sections~\ref{sec:intro_phen} and~\ref{sec:results-discussion} for further discussion of this point), so that the ratios of systematic to statistical errors quoted above are also applicable to the respective $\ell_i^{V_1^0 V_2^0 V_3^0}  m_Z^2$ results reported in the table.

A full description of the treatment of systematic effects in the channels contributing to \ZZ\ production has been given in~\cite{zz}. 
In the \qqqq\ channel, the dominant effect arises from uncertainties in the modelling of the main source of background, namely production of the $q {\bar q} (\gamma)$ final state, when the subsequent hadronization of the quarks gives rise to several jets. In the present study, the effect of this background was estimated by assuming an uncertainty of $\pm 5\%$ in the $q {\bar q} (\gamma)$ production cross-section. In the \qqll\ channel, the dominant systematic effect relevant to the present study comes from the uncertainty in the efficiency for selecting \qqee\ and \qqmm\ events, taken to be $\pm 3\%$. In addition, in the \qqee\ channel a relative uncertainty of $\pm$15\% was estimated in the calculation of the background level. In the \qqvv\ channel, as in \qqqq, the main source of systematic error arises from modelling of the \qqbrg\ background, in this case corresponding to the kinematic region with large missing energy, and hence low visible $q {\bar q}$ energy. A study of the energy flow in this region using events at the $Z$ peak allowed a determination of the effect of this uncertainty in the present analysis; it gives rise to systematic errors in the coupling parameters of order 5\% - 10\% of the values of the statistical errors. Another, comparable source of systematic error in this channel comes from the uncertainties in the cross-sections for the dominant background channels, particularly \Wev\ production. Systematic effects in the \llvv\ channels were found to be negligible. In addition, the effects of uncertainties of $\pm 2\%$ in the overall \ZZ\ cross-section and of $\pm 1\%$ in the luminosity measurement were considered. The combined effect of all the systematic uncertainties in the channels contributing to \ZZ\ production is small, typically $\sim 15\%$ of the statistical errors, and, as in the case of the $h_i^V$-related parameters discussed above, this ratio of systematic to statistical effects is also applicable to the results for the $\ell_i^{V_1^0 V_2^0 V_3^0}  m_Z^2$ related to on-shell $f_i^V$ parameters shown in table~\ref{table:results}a).

The systematic uncertainties in the study of the \qqee\ and \qqmm\ channels contributing to \Zgst\ production have been described in~\cite{zgst}. Several effects, including uncertainties in lepton identification, the effect of limited simulated data and, in the \qqee\ channel, identification of fake electrons coming from background channels, combine to give a systematic error on the efficiency to select \qqee\ and \qqmm\ events of $\pm 5\%$ and a relative uncertainty in the background level of $\pm 15\%$. In addition, a systematic error of $\pm 1\%$ in the luminosity measurement was assumed. The overall effect of these systematic uncertainties in the determination of the coupling parameters is small in comparison with the statistical errors. In combination with \ZZ\ data to produce the results for parameters $\ell_2^{\Zgg}$ and $\tilde{\ell}_4^{\ZZg}$ listed in table~\ref{table:results}b), they amount to $\sim$15\% and $\sim$5\% of the statistical errors, respectively.

In the combination of data from the different final states, \Zg, \ZZ\ and \Zgst, all the systematic effects listed above were treated as uncorrelated except those arising from the uncertainty in the luminosity measurement.

\subsection{Discussion}
\label{sec:results-discussion}

\begin{figure}[ht]
\centerline{
          \epsfig{file=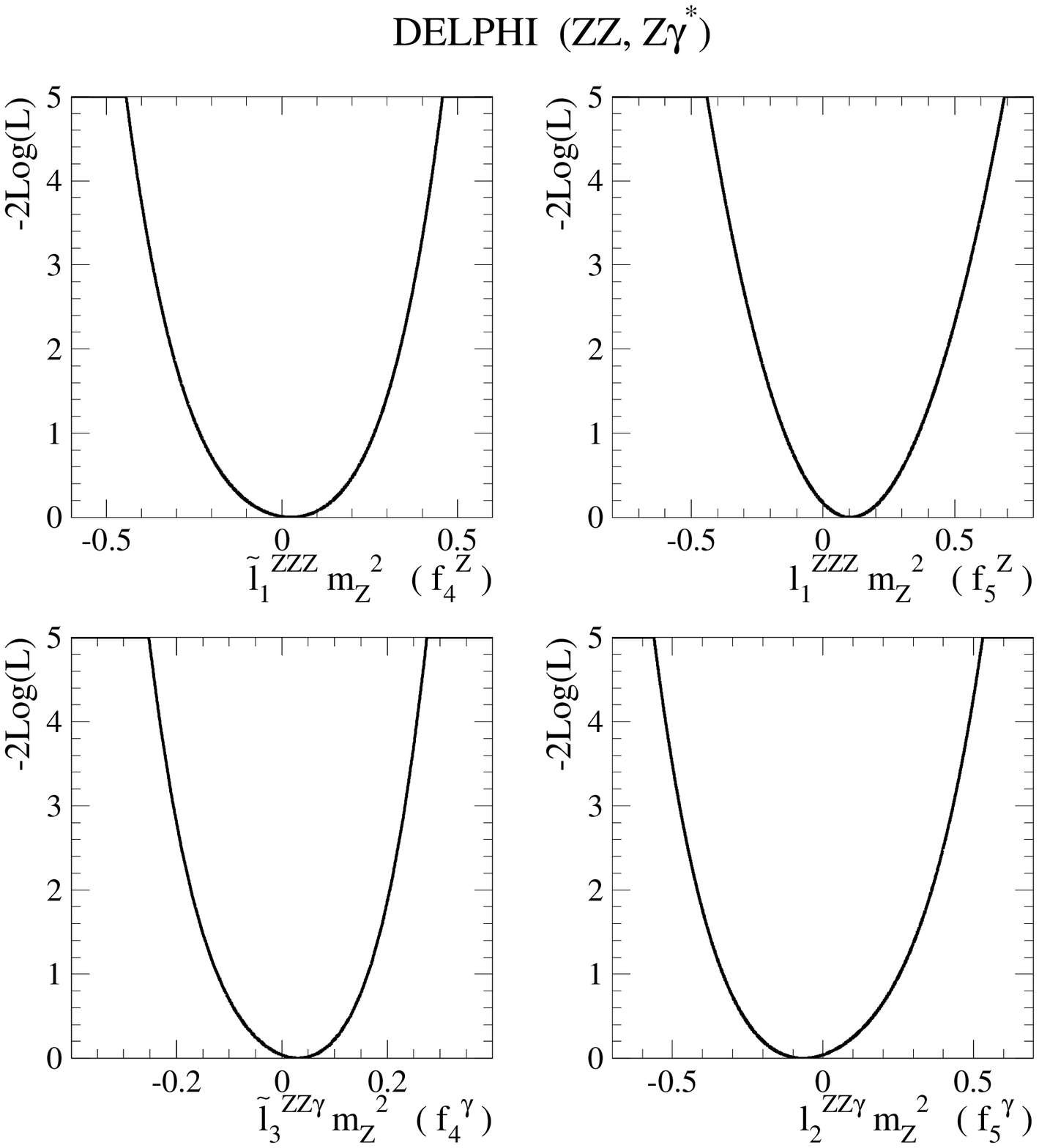,width=14cm}}
\caption{
Likelihood distributions for neutral gauge coupling parameters corresponding to Lagrangian operators influencing \ZZ\ and \Zgst\ production. The parameters are defined in section~\ref{sec:intro_phen}; the corresponding on-shell parameters are shown in parentheses on the abscissa labels. The distributions include the  contributions from both statistical and systematic effects. 
}
\label{fig:fi}
\end{figure}

\begin{figure}[ht]
\centerline{
          \epsfig{file=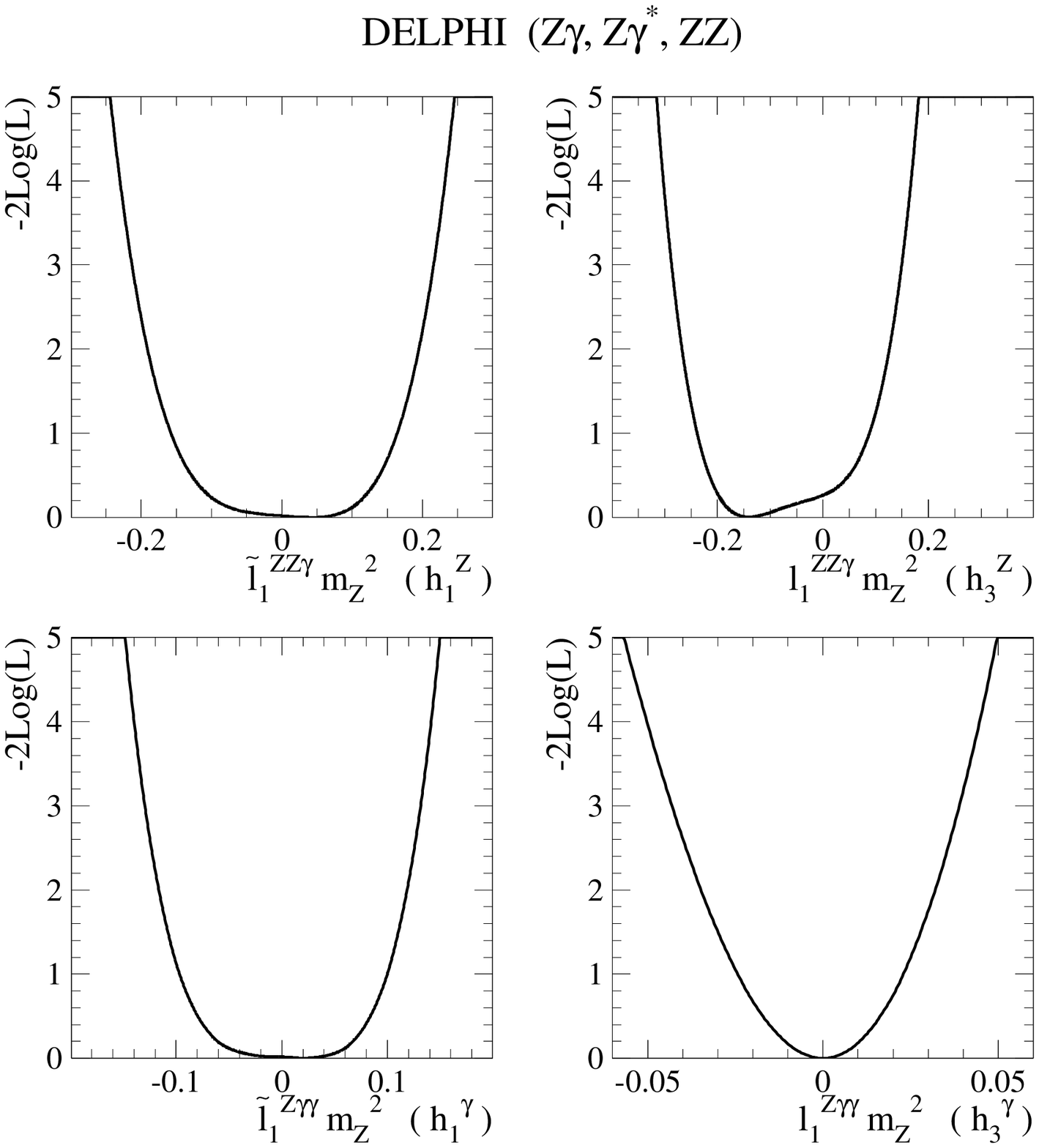,width=14cm}}
\caption{
Likelihood distributions for neutral gauge coupling parameters corresponding to Lagrangian operators influencing \Zg, \Zgst\ and \ZZ\ production. The parameters are defined in section~\ref{sec:intro_phen}; the corresponding on-shell parameters are shown in parentheses on the abscissa labels. The distributions include the  contributions from both statistical and systematic effects.
}
\label{fig:hi} 
\end{figure}

\begin{figure}[ht]
\centerline{
          \epsfig{file=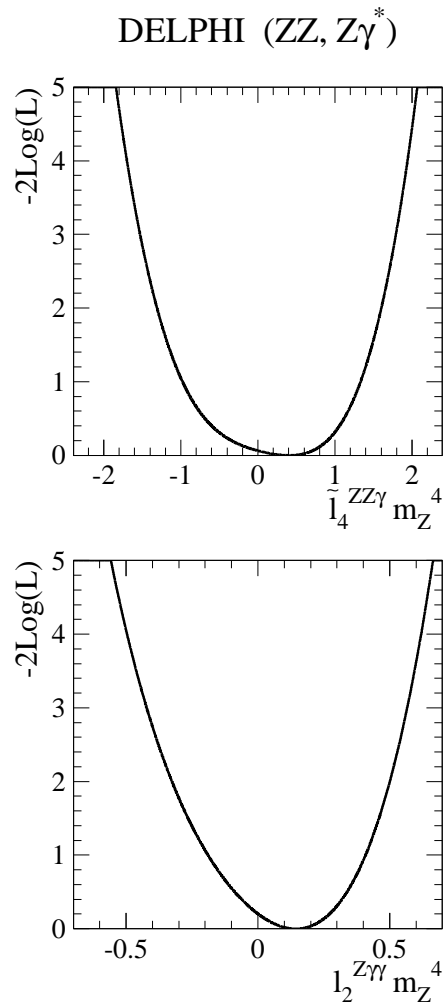,width=14cm}}
\caption{
Likelihood distributions for neutral gauge coupling parameters corresponding to Lagrangian operators affecting only the $V^0 Z \gamma^*$ vertices. The parameters are defined in section~\ref{sec:intro_phen}. The distributions include the contributions from both statistical and systematic effects. 
}
\label{fig:li}
\end{figure}

\begin{figure}[ht]
\centerline{
          \epsfig{file=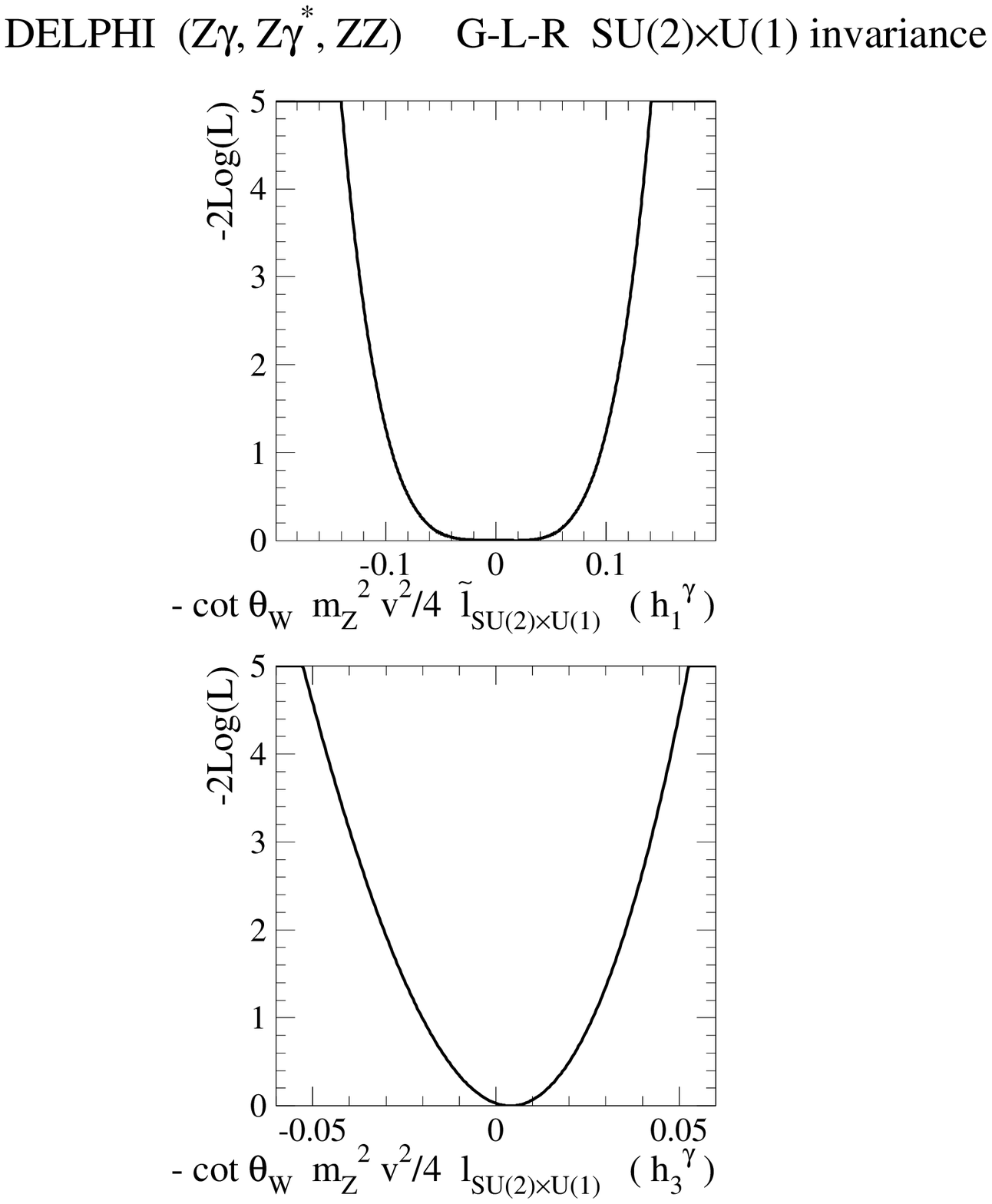,width=14cm}}
\caption{
Likelihood distributions for neutral gauge coupling parameters corresponding to $SU(2) \times U(1)$-conserving Lagrangian operators satisfying the Gounaris-Layssac-Renard (G-L-R) constraints. The parameters are defined in section~\ref{sec:intro_phen}. The distributions include the  contributions from both statistical and systematic effects. 
}
\label{fig:gounaris}
\end{figure}

\begin{figure}[ht]
\centerline{
          \epsfig{file=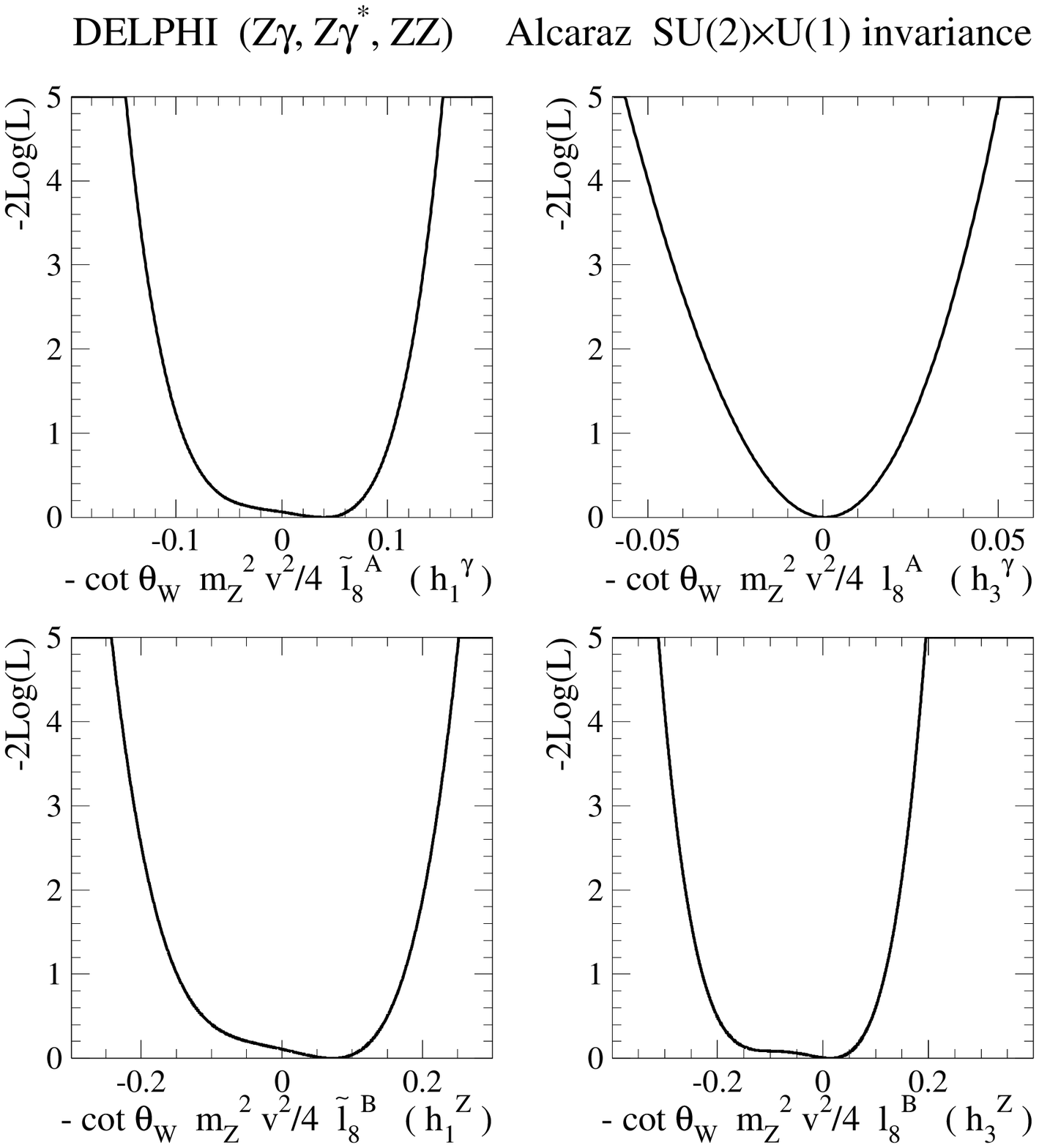,width=14cm}}
\caption{
Likelihood distributions for neutral gauge coupling parameters corresponding to $SU(2) \times U(1)$-conserving Lagrangian operators satisfying the Alcaraz constraints. The parameters are defined in section~\ref{sec:intro_phen}. The distributions include the  contributions from both statistical and systematic effects. 
}
\label{fig:alcaraz}
\end{figure}

A few comments may be made on the results shown in table~\ref{table:results} and figures~\ref{fig:fi}-\ref{fig:alcaraz}. 

All the results are compatible with the Standard Model expectation of the absence of neutral triple-gauge-boson couplings. The results shown in table~\ref{table:results}a) and figures~\ref{fig:fi} and~\ref{fig:hi} demonstrate this conclusion in the effective Lagrangian model of reference~\cite{renard} for the $d=6$ operators which, in the on-shell limit, contribute either to \ZZ\ or to \Zg\ production. As mentioned in sections~\ref{sec:intro_phen} and~\ref{sec:systematics} (and predicted from studies of simulated events~\cite{alcaraz}), the contribution to these results of the off-shell data included in their determination is small: using only the off-shell data leads to precisions poorer by factors of $\sim 3 - 7$ than using the on-shell \Zg\ or \ZZ\ data. (This effect is observed most strongly in the case of the determination of $h_3^\gamma$, where  the interference in the squared matrix element between the anomalous and Standard Model amplitudes leads to a relatively precise determination of this parameter). Thus these results, with negligible changes, may also be interpreted in terms of the parameters $h_i^V$ and $f_i^V$ of on-shell \Zg\ and \ZZ\ production, listed in the right-hand column of the table, and they may be compared directly with other published results for these on-shell parameters.

The results shown in table~\ref{table:results}b) and figure~\ref{fig:li} examine the possibility of four-fermion production via an anomalous $V^0 Z \gamma^*$ vertex by determining the coefficients of the lowest dimension ($d=8$) operators in the model of reference~\cite{renard} which would contribute to such a process. As noted in section~\ref{sec:intro_phen}, contributions from these operators affect both the \Zgst\ and \ZZ\ final states; in the determination of $\ell_2^{\Zgg}$,  the experimental samples from the two final states contribute roughly equally to the log likelihood distribution in the combination of data, while in the determination of $\tilde{\ell}_4^{\ZZg}$ the \Zgst\ contribution dominates. The results of the fits show that there is no evidence in the data for a $CP$-conserving anomalous coupling at the $\gamma Z \gamma^*$ vertex or for a $CP$-violating coupling at the $ Z Z \gamma^*$ vertex.  

The results shown in table~\ref{table:results}c) and figures~\ref{fig:gounaris} and~\ref{fig:alcaraz} indicate that there is no evidence in the data for $SU(2) \times U(1)$-conserving anomalous couplings in the models of references~\cite{renard} and~\cite{alcaraz}. Here again, in the combinations of data from different final states, the contributions from \Zg\ production dominate, as can be seen by comparison of the likelihood curves of figure~\ref{fig:hi} and either figure~\ref{fig:gounaris} or figure~\ref{fig:alcaraz}, and from the confidence limits shown in the table. This arises both because of the sensitivity to $h_3^\gamma$ noted above and because of the greater statistical contribution from \Zg\ compared to that from \ZZ\ production at LEP2 energies.

\section{Conclusions}
\label{sec:conclusions}

A study has been performed of the neutral triple-gauge-boson vertex using DELPHI data from the final states \Zg, \ZZ\ and \Zgst\ produced at LEP2. The results have been interpreted in terms of various models of the interaction Lagrangian proposed in the literature. We find no evidence for the production of these states by processes involving neutral triple-gauge-boson vertices with either one or two off-shell bosons, nor when the data are analyzed in terms of models in which the neutral 
triple-gauge-boson vertex is constrained to be $SU(2) \times U(1)$-conserving. These conclusions are in agreement with the predictions of the Standard Model.

%\input{acknow.tex}
%\input{acknowledgements}
%         Modified on 04-06-1999 by dimartino
%-------------------------------------------------------------------
\subsection*{Acknowledgements}
\vskip 3 mm
We are greatly indebted to our technical 
collaborators, to the members of the CERN-SL Division for the excellent 
performance of the LEP collider, and to the funding agencies for their
support in building and operating the DELPHI detector.\\
We acknowledge in particular the support of \\
Austrian Federal Ministry of Education, Science and Culture,
GZ 616.364/2-III/2a/98, \\
FNRS--FWO, Flanders Institute to encourage scientific and technological 
research in the industry (IWT) and Belgian Federal Office for Scientific, 
Technical and Cultural affairs (OSTC), Belgium, \\
FINEP, CNPq, CAPES, FUJB and FAPERJ, Brazil, \\
%Czech Ministry of Industry and Trade, GA CR 202/99/1362,\\
%Ministry of Education of the Czech Republic LA134,\\
Ministry of Education of the Czech Republic, project LC527, \\
Academy of Sciences of the Czech Republic, project AV0Z10100502, \\
Commission of the European Communities (DG XII), \\
Direction des Sciences de la Mati$\grave{\mbox{\rm e}}$re, CEA, France, \\
Bundesministerium f$\ddot{\mbox{\rm u}}$r Bildung, Wissenschaft, Forschung 
und Technologie, Germany,\\
General Secretariat for Research and Technology, Greece, \\
National Science Foundation (NWO) and Foundation for Research on Matter (FOM),
The Netherlands, \\
Norwegian Research Council,  \\
State Committee for Scientific Research, Poland, SPUB-M/CERN/PO3/DZ296/2000,
SPUB-M/CERN/PO3/DZ297/2000, 2P03B 104 19 and 2P03B 69 23(2002-2004)\\
FCT - Funda\c{c}\~ao para a Ci\^encia e Tecnologia, Portugal, \\
Vedecka grantova agentura MS SR, Slovakia, Nr. 95/5195/134, \\
Ministry of Science and Technology of the Republic of Slovenia, \\
CICYT, Spain, AEN99-0950 and AEN99-0761,  \\
The Swedish Research Council,      \\
Particle Physics and Astronomy Research Council, UK, \\
Department of Energy, USA, DE-FG02-01ER41155, \\
EEC RTN contract HPRN-CT-00292-2002. \\

%=========================================================================%

\newpage
%=========================================================================%

%%%%%%%%%%%%%%%%%%%%%%%%%%%%%%%%%%%%%%%%%%%%%%%%%%%%%%%%%%%%%%%%%%%%%%%%%%
%BIBLIOGRAPHY
%%%%%%%%%%%%%%%%%%%%%%%%%%%%%%%%%%%%%%%%%%%%%%%%%%%%%%%%%%%%%%%%%%%%%%%%%%

%\newpage
\clearpage

\begin{thebibliography}{99}
%\bibliographystyle{unsrt}
\newcommand{\bi}{\bibitem}
%\itemsep -5pt
%\parsep -5pt

\bi{hagiwara} K.~Hagiwara {\it et al.}, Nucl. Phys. {\bf B282} (1987) 253.

\bi{renard} G.J.~Gounaris, J.~Layssac and F.M.~Renard, Phys. Rev. {\bf D62} (2000)  073012.

\bi{bilenky} M.S.~Bilenky {\it et al.}, Nucl. Phys. {\bf B409} (1993) 22.

\bi{yb} G.~Gounaris {\it et al.}, in  {\it Physics at
        LEP2}, eds. G.~Altarelli, T.~Sj\"ostrand and F.~Zwirner, CERN 96-01 (1996) Vol.1, 525.

\bi{renard-addendum}  G.J.~Gounaris, J.~Layssac and F.M.~Renard, addendum to~\cite{renard} above, in hep-ph/0005269 (2000).

\bi{alcaraz} J.~Alcaraz, Phys. Rev. {\bf D65} (2002) 075020.
 
\bi{zg}  DELPHI Collaboration, W.~Adam  {\it et al.}, Phys. Lett. {\bf B380} (1996) 471; \\
 DELPHI Collaboration, P.~Abreu  {\it et al.}, Phys. Lett. {\bf B423} (1998) 194.

\bi{zz} DELPHI Collaboration, J.~Abdallah {\it et al.}, Eur. Phys. J. {\bf C30} (2003) 447.

\bi{zgst} DELPHI Collaboration,  J.~Abdallah {\it et al.}, {\it Z}$\gamma^\ast$ {\it production in \ee\ interactions at $\sqrt{s}$ = 183 - 209~GeV}, 
Accepted by Eur. Phys. J. C, arXiv:0706.2565

%\bi{zgst} DELPHI Collaboration,  M.~Begalli, E.~Graziani and M.E.~Pol,  {\it Z$\gamma^\ast$ %production in \ee\ interactions at $\sqrt{s}$ = 183 - 209~GeV}, submitted to ICHEP~2004, Beijing %(2004).

\bi{opal1} OPAL Collaboration G.~Abbiendi {\it et al.},  Eur. Phys. J. {\bf C17} (2000) 553.

\bi{lep_ntgc} ALEPH Collaboration, ALEPH 2001-061 CONF 2001-041 (2001);\\
L3 Collaboration, P.~Achard {\it et al.},  Phys. Lett. {\bf B572} (2003) 133; \\
L3 Collaboration, P.~Achard {\it et al.},  Phys. Lett. {\bf B597} (2004) 119;  \\
OPAL Collaboration, G.~Abbiendi {\it et al.}, Eur. Phys. J. {\bf C32} (2004) 303.

\bi{delphi_det}  DELPHI Collaboration, P.~Aarnio {\it et al.}, Nucl. Instr. and Meth. {\bf A303} (1991) 233.

\bi{delphi_perf} DELPHI Collaboration, P.~Abreu {\it et al.}, Nucl. Instr. and Meth. {\bf A378} (1996) 57.

\bi{trigger} DELPHI Trigger Group, A.~Augustinus {\it et al.}, Nucl. Instr. and Meth. {\bf A515} (2003) 782.

\bi{delphi_det2} DELPHI Silicon Tracker Group, P.~Chochula {\it et al.}, Nucl. Instr. and Meth. {\bf A412} (1998) 304.

\bi{gx} DELPHI Collaboration, J.~Abdallah {\it et al.},  Eur. Phys. J. {\bf C38} (2005) 395.

\bi{nicro1} G.~Montagna {\it et al.}, Nucl. Phys. {\bf B452} (1995) 161.

\bi{koralz}  S.~Jadach,  B.F.L.~Ward and Z.~Was, Comp. Phys. Comm. {\bf 79} (1994) 503.

\bi{sprime} P.~Abreu {\it et al.}, Nucl. Instr. and Meth. {\bf A427} (1999) 487.

\bi{pythia}  T.Sj\"{o}strand, {\it PYTHIA 5.7 / JETSET 7.4}, CERN-TH 7112/93 (1993).

\bi{baur} U.~Baur and E.L.~Berger, Phys. Rev. {\bf D47} (1993) 4889.

\bi{ifact} G.J.~Gounaris, J.~Layssac and F.M.~Renard, Phys. Rev. {\bf D61} (2000)  073013.

\bi{ida} T.G.M.~Malmgren, Comp. Phys. Comm. {\bf 106} (1997) 230;\\
            T.G.M.~Malmgren and K.E.~Johansson, Nucl. Instr. and Meth. {\bf 403} (1998) 481.

\bi{excalibur} F.A.~Berends, R.~Pittau and  R.~Kleiss, Comp. Phys. Comm. {\bf 85} (1995) 437.

\bi{grc4f} J.~Fujimoto {\it et al.}, Comp. Phys. Comm. {\bf100} (1997) 128.

\bi{bhwide} S.~Jadach, W.~Placzek and B.F.L.~Ward, Phys. Lett. {\bf B390} (1997) 298. 

\bi{twogam} T.~Alderweireld {\it et al.},  in {\it Reports of the Working Groups on Precision Calculations for LEP2 Physics}, eds. S.~Jadach, G.~Passarino and R.~Pittau,  CERN 2000-009 (2000) 219.

\bi{bdk} F.A.~Berends, P.H.~Daverveldt and R.~Kleiss, Comp. Phys. Comm. {\bf 40} (1986) 271, 285 and 309. 

\bi{deltgc}  O.P.~Yushchenko and V.V.~Kostyukhin, {\it DELTGC - A program for four-fermion calculations},  DELPHI 99-4 PHYS 816 (1999).

\bi{wphact}
E.~Accomando and A.~Ballestrero, Comp. Phys. Comm. {\bf 99} (1997) 270; \\
E.~Accomando, A.~Ballestrero and E.~Maina, Comp. Phys. Comm. {\bf 150} (2003) 166; \\
A.~Ballestrero, R.~Chierici, F.~Cossutti and E.~Migliore, Comp. Phys. Comm. {\bf 152} (2003) 175.

\bi{kk2f}
S.~Jadach, B.F.L.~Ward and Z.~Was, Comp. Phys. Comm. {\bf 130} (2000) 260. 

\bi{herwig} G.~Marchesini {\it et al.},  Comp. Phys. Comm. {\bf 67} (1992) 465.

\end{thebibliography}
\end{document}